# Open clusters: time-scales, core collapse and blue stragglers

Félix Llorente de Andrés[1,2,*], Carmen Morales- Durán[1,†]

[1]Departamento de Astrofísica, Centro de Astrobiología, European Space Astronomy Center (ESAC) Campus, Villanueva de la Cañada (Madrid), Spain.

[2]Ateneo de Almagro, Sección de Ciencia y Tecnología, Almagro (Ciudad Real), Spain.

[†]This work is dedicated to the memory of Dr. Morales- Durán who died on February 15, 2022.

Email adress:

fllorente@cab.inta-csic.es

## Abstract

We developed a mathematical model to derive time scales and the presence of BS stars. The model is based on the variation of mass through a circle into the cluster defined by a radius, and at a time; this mass cross is translated into a differential equation that it can be integrated for a given radius (r) and a determined time (t). From this equation we can derive the different time scales that allows us to reach conclusions like: clusters not containing blue strugglers (BS) stars disappear younger than those clusters containing BS. In clusters containing BS stars, the volume which takes up half of the cluster mass is bigger than the one corresponding to clusters without BS stars but the time to catch it up is shorter. We also studied, by means of this equation, the core collapse of stars of the cluster and the region where this concentration is stopped/retained; this region is identified by means of the relation c/ch, being c=log(rt/rc) and ch=log(rc/rh). Where rt and rc are the tidal and the core radius respectively, and rh is the radius where half of the cluster mass is concentrated. The model also drove us to the conclusion that the number of the blue straggler stars in a cluster follows a distribution function whose components are the ratio between relaxation time and the age, labelled as ƒ, and a factor, named ϖ, which is an indicator of the origin of the BS; ϖ increases as the number of BS increase but it is limited to ~5.0. The mentioned distribution function is expressed as $NBS \sim f^3(\frac{1}{e^{f/\varpi}-1})$. The validity of this function was carried out by means of matching the number of observed blue straggler (BS) stars to the number of predicted ones in the available sample of OC.

## 1. Introduction

Before becoming a cluster the interstellar clouds maintain their mechanical equilibrium through magnetic fields, turbulence, and rotation (Shu et al. 1987). After formation, all clusters suffer significant infant mass loss, while a large fraction undergoes infant mortality. At this point, the formation of an open cluster will depend on whether the newly formed stars are gravitationally bound to each other; otherwise an unbound stellar association will result. Moreover, interactions with molecular clouds or tidal stripping by the gravitational field of their host galaxy tend to dissolve or destroy the formed cluster. All these mechanisms effectively remove 50-90% of the clusters that are formed in a galaxy, within the first 10Myr of their lives (Lada & Lada 2003, Fall 2004; Bastian et al. 2005 and de Grijs 2010).

On the other hand, 80% of the old open clusters are farther away than the Sun from the Galactic Centre. This clearly indicates that the survival times of open clusters (OCs) at smaller Galactocentric distances from the Sun are significantly shortened due to the stronger tidal forces and the more frequent encounters with giant molecular clouds (GMCs) (Foebrich et al., 2010).

Clusters with enough mass to be gravitationally bound and, once the surrounding nebula has evaporated, can remain for many tens of millions of years, but internal and external processes tend, along the time, to disperse them, in a gradual 'evaporation' of cluster members. Internally stellar radiation, stellar evolution, stellar close encounters etc. can impel an energy which is translated to the stars as much as to get a speed beyond the escape velocity of the cluster. Externally, clusters tend to be disturbed by external factors such as passing close to or through a molecular cloud.

This combination of different effects internal and external (stellar evolution, encounters with GMCs, tidal stripping, etc.) yields in decreasing mass of the cluster until the cluster is diluted. Bastian et al. (2013) concluded that, for clusters, the time-scale of their dilution depends on the initial conditions of the cluster: the stellar initial mass function, its concentration and the tidal forces suffered by the cluster during its galactic orbit. These tidal shocks facilitate the dilution of the cluster via stars escaping from the cluster but, in principle such stars would be the less massive ones; in lesser extend the blue straggler (BS) stars because they are commonly accepted as the most massive stars and sink to the cluster centre. Chen and Han (2009) argued that there is a process which brings new hydrogen into the core and therefore "rejuvenate" a star remain longer in the MS stage; this hydrogen is transported by mass-transfer in close binaries, which can have a primordial or collisional origin. This is the most accepted origin for the BS stars.

It is reasonable to think that stars actually collide within the densest stellar systems but rarely open clusters; however, being rare, this phenomenon could be more common than expected (Glebbeck and Pols, 2008). But it is also reasonable to think that encounters are not enough as to create the number of observed BS stars by merging; the rest of them must be due to the presence of primordial binaries. Such encounters can take place in open clusters depending on several factors, moreover a combination of factors like concentration, relative density and cluster time-scale can provide, jointly with relaxation, crossing, encounter timescale, etc. the conditions for strong encounters to happen, creating merger stars which are added to the previous existence of primordial binaries; whose number is much larger than stars merged by collisions.

However, none of the studies, as far as we know, have dealt with the structural and dynamic difference between clusters that contain BS and those that do not. As well as an explanation of the number of observed BS that contains the cluster.

The aim of the present work is to identify the factors which would play an important role into the game of BS stars versus the cluster dynamics. In order to accomplish with this aim we have built two sample of clusters, with and without BS stars; the first on of them (190 OCs) was taken from Ahumada and Lapasset (2007), where we performed our complete test of our model and second one from de Marchi et al (2010) for corroborating results of presence of BS stars, obtained previously with the first set of OC. The OCs characteristics were quoted from Kharchenko et al (2013).

## 2. Methodology, Models and Results

### 2.1. *Time-scales.*

A combination of different effects: stellar evolution, encounters with GMCs, tidal stripping, etc. yields in decreasing mass of the cluster until …. An open question arises: will the cluster be completely destroyed or diluted? It seems that the answer is NO completely destroyed but, it could be dispersed, maintaining a hard core. To reply this question we have to investigate the dilution of the cluster due to stars escaping from the outer shells of the cluster, as well as the concentration of stars in a dense core by sinking the more massive stars to the cluster centre. We try to find a plausible answer, by means of developing a mathematical model which allows us to compute the different time-scales, basis of our study.

Starting from the hypothesis that the cluster is inside a sphere of radius r and that due to the phenomena mentioned above it is going to expand in such a way that we can assume that the mass variation that crosses the edge can be expressed by dM /dt which is equal to:

$$\frac{dM}{dt} = \frac{d(\rho V)}{dt} = 4\pi r^2 \rho \left(\frac{dr}{dt}\right) + \frac{4}{3}\pi r^3 \left(\frac{d\rho}{dt}\right) \qquad \text{Equ. (1)}$$

V represents the volume and ρ the density.

It could be assumed that the distribution of stars inner to the cluster is conserved mainly due to three parameters, related to the star motion: energy, angular momentum and energy of star's motion perpendicular to the galactic plane. But the decisive parameter is the energy (Danilov, 2002a and 2002b). Recently Trujillo-Gómez et al (2019) showed that the minimum and maximum cluster mass are equal because the entire cloud mass spectrum will collapse into a single bound object at the maximum mass scale

We are able to assume that a star is kept within the host cluster if its kinetic energy is, at least, equal to its potential energy derived from the gravitational attraction which can be expressed in terms of speed; v = ( GM/r) ½. Thus Equ 1 can be written as:

$$\frac{dM}{dt} = \frac{d(\rho V)}{dt} = 4\pi r^2 \rho \left(\frac{GM}{r}\right)^{1/2} + \frac{4}{3}\pi r^3 \left(\frac{d\rho}{dt}\right)$$

According to Gieles et al (2010) whether the cluster stays in the same orbit there is no dynamical friction, consequently the mean density will remain almost constant. There would be some sub tilts but second order effects (Gieles et al. 2014), in fact the initial cluster mass is almost the same as it is when the cluster dissolution; in other words ρ could remain actually constant until the cluster starts to evaporate. This assumption is also supported by the work performed by Lamers et al (2010). Their argument is based on the tidal radius (rt) as a function of the cluster mass (Mcl) and the radius of the galaxy (Rgal) in terms of Galaxy mass (Mgal) within the cluster orbit, i.e. $rt^3 \sim Mcl(Rgal^3)/Mgal$. If Rgal and Mgal do not change (cluster stay in the same orbit; no dynamical friction) thus the ratio $(Rgal^3)/Mgal$ will remain the same and so $rt^3 \sim Mcl$. Thus equation (1) becomes:

$$\chi^{-\frac{1}{2}} \left(\frac{d\chi}{dt}\right) \sim K\, r^{3/2} \qquad \text{Equ. (2)}$$

Where, χ= Mass, and K = $(4\pi\rho G)^{1/2}$ is a constant. Then, the solution of equation (2) is:

$$\chi \sim r^3\, t^2 + intercept \qquad \text{Equ. (3)}$$

Let us call χ as "function of mass", in fact is the mass, within a volume of radius, r, and at any corresponding time t. At t =0 the gravitational attraction to form the cluster appears when a minimum value of the mass collapse Trujillo-Gómez et al (2019).

Equation 3 is in agreement with various previous works, mainly, Lamers et al (2005) and references therein. Lamers et al. (2010) indicated that in several papers it is showed that the *time-scale* $\alpha\, Mcl^{\gamma}$ (Mcl is the cluster mass). Observations agree with γ = 0.62 predicted for tidal effects. From our equation we derived that $time-scale\ \alpha\ \chi^{0.5}$, in our sample of OCs. Remark that in the equation (3), time is also present because even unbound stars need some time to leave completely its cluster.

Assuming that the cluster is start to evaporated when r_rt ) to the tidal radius), we can write:

$$log\chi r - log\chi rt \sim 2\,(log\ age - log\ teva)$$

Thus evaporation time, teva, can be computed by:

$$log\ teva \sim log\ age + (log\chi rt - log\chi r)/2 \qquad \text{Equ. (4)}$$

Our sample of clusters, as mentioned in the introduction, is taken from the AL 07 catalogue and their respective cluster characteristics from K13; so that N (total number of stars) would correspond to N1sr2, r (cluster radius) which corresponds to r2 (but in terms of linear radius), rt and rc correspond to tidal radius and core radius respectively.

To compute the (4) it should be necessary to make an abstraction: the cluster does not start to evaporate before r reach rt. Thus we computed the value of (logχ_rt-logχ_r)/2, called hereinafter Δ, for our cluster sample being the respective age that published by K13. The value of Δ varies between 0.43 and 1.20 for clusters with BS stars and between 0.48 and 1.06 (exceptionally the maximum value is 1.79 for the cluster NGC2264 - very young cluster) for clusters without BS stars. What is translated to: clusters with BS stars exist till log age ~10.5; clusters not containing BS stars exist till log age ~9.5. Consequently, clusters not containing BS stars disappear younger than those clusters containing BS. We have also reproduced these same calculi taking the age from AL07 (we also checked that ages from AL07 are those ones listed in the WEBDA catalogue). We did not find difference at all in respect with the values of Δ neither for cluster with BS nor with cluster without BS. We noted that there is a slight difference between the two age scales: log age = 10.8 for cluster with BS and and log age = 9.2 for cluster without BS. In fact the correlation coefficient between both scales of ages – K13 and AL07 - is ~ 0.895. However there are four clusters (IC 5146, NGC 1444, Berkeley 79 and NGC 2353; which contain only one BS star exception made of NGC144 with two BS stars) listed as young open clusters from K13 but they are significantly older.

The same way was followed for computing log χ corresponding to rc; then we can derive the respective time-scales following the way followed in computing log teva.

The correlation between the three log χ (R, rt and rc) is linear with a goodness of fit which indicates the homogeneity of equation (3). This led us to use the same method for computing log χrh, being rh the radius where half of the mass is concentrated.

According to our previous definition of χ as a "function of mass", it varies as $r^3$. In addition a relationship exists between log χ and log N as a function of time. However in our work such a correlation – see figure 1 - is moving into two limits: a maximum of log χ R ~ 22 and a minimum log χ R~ 12 independently of the total mass of the cluster.

It is clear, even obvious, that the evaporation is accelerated by tidal shocks which propel additional energy to the stars, thus they can escape. However figure 2 shows that most of the clusters containing BS stars are within an interval of the concentration c, defined as c = log (rt/rc), between 0.4 and 1; that means that rt is in between 2.5 and 10 times the core radius. In addition from figure 2 as a conclusion we might say: in general clusters containing BS stars live longer than those ones without BS stars and their concentration limits are also lower.

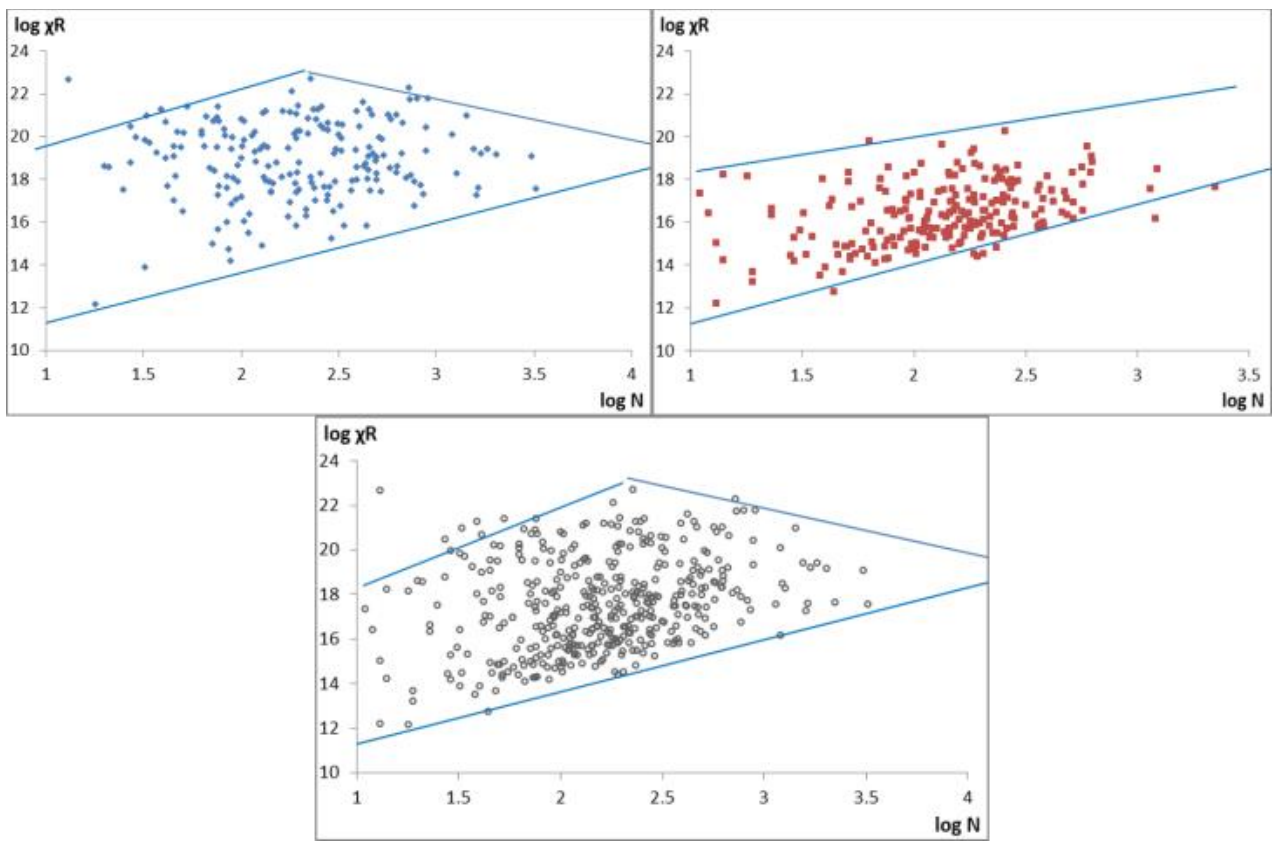

*Figure 1. The two lines represent the two limits of log χR; with the exception of Berkeley 18 and Berkeley 20.*
*Note for figures: the key of colors and geometry of data will be the same along the present work: blue diamonds, red squares and open points correspond, respectively, to clusters containing BS stars, clusters not containing BS stars and altogether without distinction.*

Let us follow similar assumptions than those we adopted when the cluster is diluting but actually is also sinking down, in this case R is trending to become rc ( R$\to$ rc).

Operating as we did before; the equation (4) could represent the time of collapse, tcol:

$$\log tcol = \log age + (log\chi R - log\chi rc)/2 \quad \text{Equ (5)}$$

The value of (logχR - logχrc), of equation 5, varies between 0.47 and 1.14 for clusters without BS stars and between 0.65 and 1.34 for cluster with BS stars; consequently its evolution with time follows different ways.

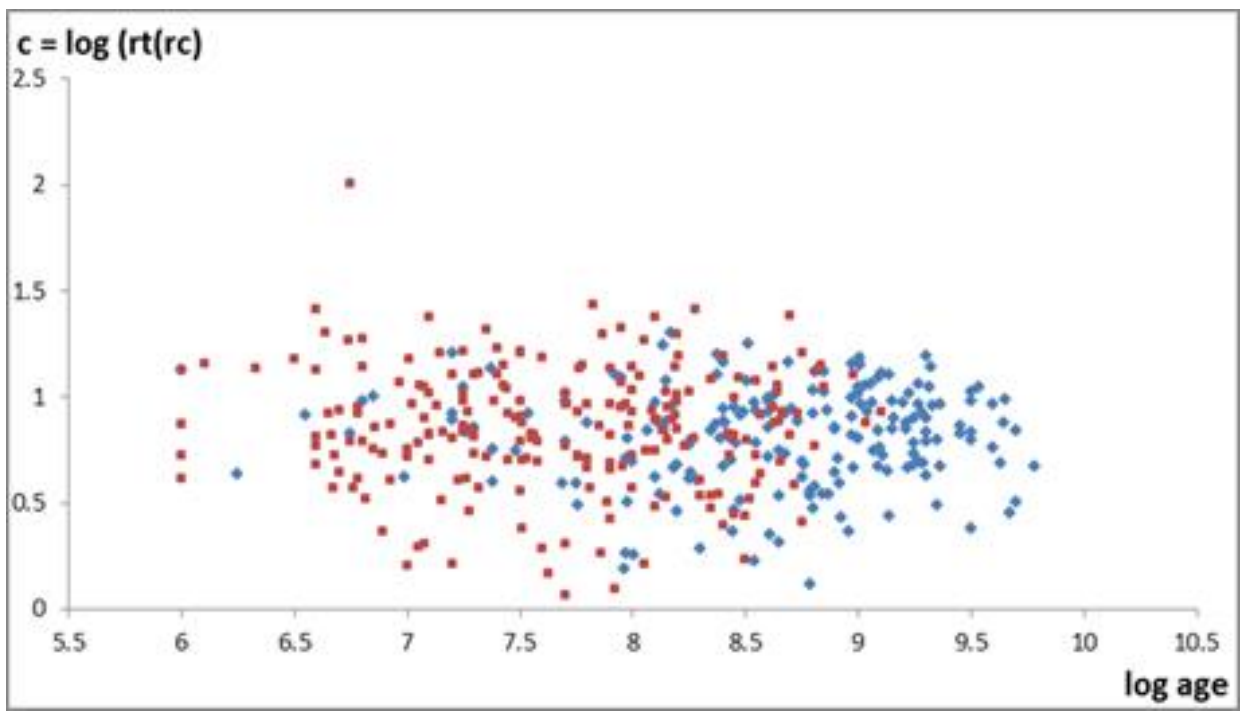

Figure 2. *It shows how the central concentration c is trending to a value 0.4 for clusters containing BS stars and displays a big scattered dispersion for cluster not containing BS stars.*

Since BS stars are more massive than normal stars, they sink toward the center quickly after formation, which implies that the BS star population should be less affected by cluster mass loss than the normal stellar population. This would explain of the Δ variation in equations (4) and (5) between cluster with and without BS stars. However equation (5) is not representative of the cluster collapse because the more massive stars tend to have smaller velocities because they are breaking – that will be explained later. Moreover, during the core collapse, if the massive stars attempt to equalize kinetic energy with stars outside the core, they lose energy and sink even faster toward the core; the more stars are sinking the more encounters happen drifting the stars from their track. The border of this process limit is called core collapse; eventually contraction is probably halted by injection of energy from binary stars and other stars placed inside the core. This border will be described forward in subsection 2.2.

During the collapse the stars change their tracks; the scale of time where the star changes its track is defined as relaxation time. Changes of tracks could force the star to encounter (encounter time) another one and both stars merge in one star or in a pair of closed binaries; some stars travel jointly together. Thus, that could be interpreted as that relaxation will force the massive stars to the central regions, and could force the lower mass stars to the periphery. Consequently, dynamical effects will remove the stars from the outskirts, mainly the low mass stars. So as the cluster loses mainly low mass stars; in consequence, as the BS stars are more massive than normal ones, the fraction of BS will "increase" proportionally due to dynamical effects.

We compared our results with those obtained by Lamers and Gieles (2006). They predicted a dissolution time as a function of the initial mass (Mi) as $tdis = 1.7(Mi/10^4 M\odot)^{0{,}67}$ Gyr. They also derived an initial mass between $1.0*10^2$ and $3*10^4$MΘ. We have derived from our model that the initial mass should be between $0.4*10^2$ and $3*10^5$MΘ. This initial mass would be, for clusters with BS stars, between $0.63*10^2$ and $3*10^5$MΘ and for those cluster without BS stars, between $0.4*10^2$ and $8*10^4$MΘ. It seems that we have one order of magnitude higher than the values they found. The reason is that they performed their work based on clusters within a distance of 600pc. If we take from our sample those clusters within the same distance, it is reduced to 43 clusters, then the results are quite similar: for those clusters with BS stars the initial mass should be between $1.0*10^2$ and $6.8*10^4$MΘ and from $0.55*10^2$ to $2*10^4$ for those without BS stars. In both studies, Lamers and Gieles (2006) and ours, it appears that the initial mass for clusters with BS is larger than the initial mass for clusters without BS.

We have shown along this work that the behavior of clusters containing BS stars is different from those ones without the presence of BS stars. Moreover it seems that the number of BS stars is connected with the evolution of the cluster.

From figure 3, we are able to conclude that the higher the presence of BS stars, the lesser the difference exists between age and evaporation, independently of the distance. In other words, older clusters, near their evaporation time, have more BS. Such a conclusion is common and well known; it corroborates the goodness of our model.

At the end, the creation of new BS stars is a basic process due to the deflection of the paths followed by stars, in turn, due to encounters with other stars and/or due to the collective gravitational field of the system at large, as showed in the next section. It is assumed that stellar encounters lead to dynamical relaxation; the time to reach it can be estimated as a typical star to change its energy by an amount equal to the mean energy.

To achieve our goal of investigating in detail the evolution of the cluster and the presence of BS stars we have built a grid of the different time-scales; listed in table 1. The system of equations to derive the different time-scales has been quoted from P. Armitage in: http://www.astro.caltech.edu/~george/ay20/Ay20-Lec15x.pdf.

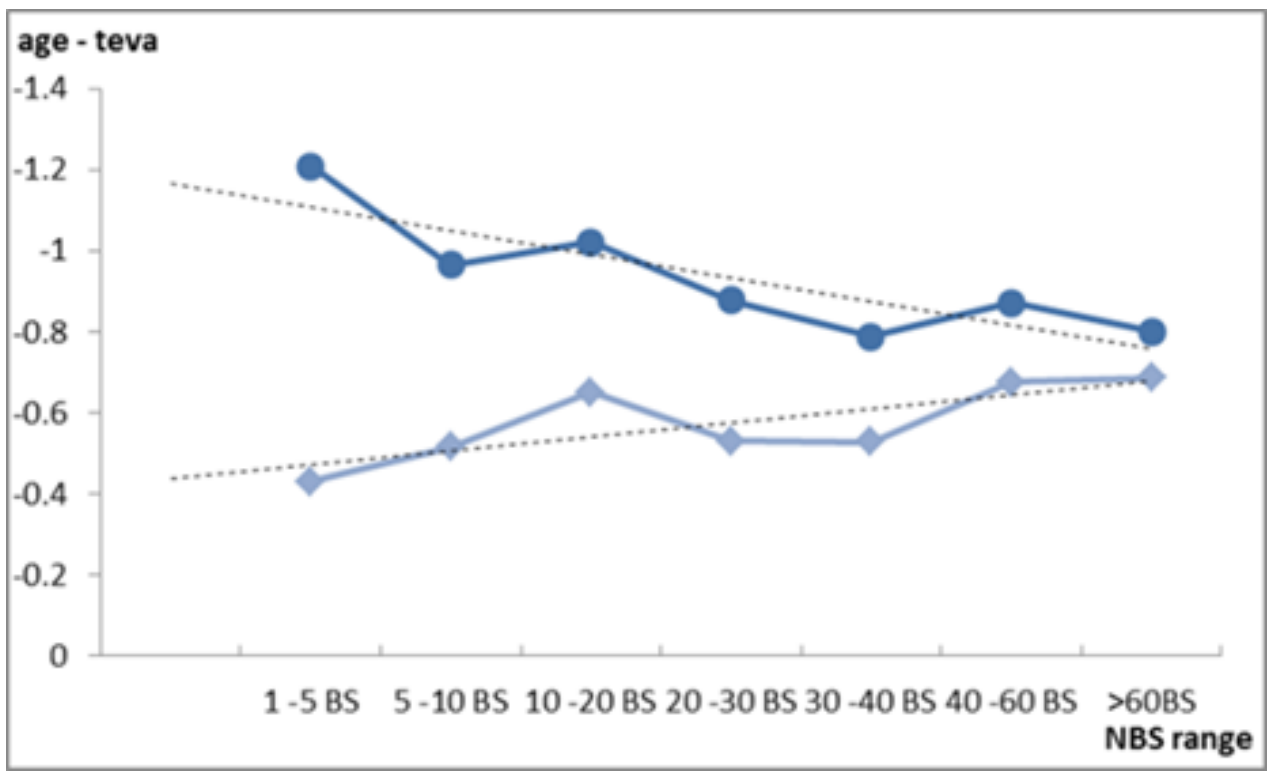

*Figure3. In this figure, is plotted a histogram built with the difference between the age of the cluster and the evaporation time vs range of BS stars. Straight line represents the tendency.*

*Table 1 Relationships among the different time-scales.*

| Equation | $R^2$ |
|---|---|
| log trlx = 0.99 log th – 0,10 | 0.98 |
| log tcrs = 0.91log th – 0,28 | 0.80 |
| log tenc = 0.99 log th +0.674 | 0.98 |
| log tenc = log trlx + 0.78 | 1.00 |
| log t rlx = 0.83 log t rc + 0.43 | 0.82 |
| log th = 0.84 log trc + 0.64 | 0.84 |

The encounter radius could be written as rec ~ $2Gm/v^2$, where m is the mass and v is the velocity. The time-scale for a star to undergo an encounter with other stars would be according to:

$$tenc = 400 * v^3 * m^{-2} * n^{-1}$$

Where v is the velocity in terms of $10 kms^{-1}$, m is the mass in M Θ and n is the density in $pc^{-3}$. This density n is the relative density. We assume that the ratio between the age of each cluster and the time-scale for a single star to undergo an encounter gives a plausible estimate of the number of encounters: $N_{enc}$ (number of encounters) = age/tenc.

The number of encounters computed for our sample of clusters younger than log age ≤8 is between 1 and 5 and the corresponding number of BS stars is between 1and 3. In between 2 to 8 for those OCs whose age is log age >8. The conclusion is that: in young OCs, BS stars come mainly from star encounters; although it is not unlikely the coexistence with primordial binaries from the very beginning of the cluster, not yet recognized as BS stars. It is also important to remember that binary stars can have fairly long lifetime before becoming BS stars, so they could have been formed long time ago when the parent stars were less evolved. However, in old clusters the number of encounters is not enough to explain the number of BS stars. In these clusters the mass transfer in binaries is probably a common accepted origin for the majority of BS stars in OCs.

Time to cross is defined as tcrs = r/v, where r is the radius of the cluster and v the corresponding speed, then:

$$\frac{trlx}{tcrs} = \frac{N}{6 * \ln(\frac{N}{2})}$$

N is the total number of stars within the cluster.

The time-scales tec, trlx and tcrs were computed in coherence with our previous estimate of teva.

In order to ensure ourselves that our computations, under the assumptions above mentioned, are within the range of values already worldwide published as average, we compared both sets of deduced time-scales. Assuming a cluster of 100 stars with a radius of about 2 pc and a crossing v of about 0.5 $kms^{-1}$, the average values of the timescales for open clusters quoted from the literature are: Crossing time log t crs = 6.3 Relaxation time log trlx = 6.9 Evaporation time log t eva = 9.0. The median values of the time-scales for our open clusters obtained from our computations are (we have adopted the median because our dispersion in the number of stars is largely spread): median Crossing time log tcrs = 6.0, median Relaxation time log trlx = 6.7, median Evaporation time log teva = 9.0. Differences in crossing time and relaxation time are due to the number of stars (our median is 165 stars) and to the dimension of the radius (our median is 2.6). These differences are minimal; in the case of evaporation time, there is not difference; this last point in is completely agreement with results published by Boutloukos and Lamers (2003). Our model also evidence that the disruption time-scale increases with the distance from the Galactic Centre (See Goodwin & Bastian (2006), Lamers & Gieles (2007) and Piskunov et al. (2007))..

From our calculations we conclude: trlx and tcrs have close similar behaviour, with a slight difference in the slopes when comparing clusters with BS stars and clusters without them; the slopes are respectively 1.0 and 0.9.

We can assert that for a smaller (larger) radius, relaxation becomes more (less) important, while the escape due to the tidal field becomes less (more) important.

#### *2.1.1. Clusters near evaporation.*

It is commonly accepted that under the conditions when tcrs ~ trlx < teva clusters quickly dissolve. In order to identify those clusters which will dissolve quickly we have selected those ones whose difference between log tcrs and log trlx is within the range ±0.08, besides that always teva > tcross.

Table 2 lists those clusters which fulfil these conditions. The galactic position is also reported. The common characteristics of the clusters are: low amount of stars, young age, not containing (or very few) BS stars, and small Galactocentric distance. They are situated near the galactic plane.

### *2.2. Concentration and core collapse*

We have previously mentioned that the concentration, c, as c=log (rt/rc) - the ratio rt/rc determines the dynamical profile of the clusters; this concentration is related to the age for two samples of clusters, with and without BS stars. So that, the figure 2 shows the ratio rt versus rc for the age range of our clusters sample, we see that the minimal concentration (log rt/rc=1.5) corresponds to tidal radius approximately equal to 30 times core radius, and the densest concentration is when tidal radius is almost equal to core radius. No clusters exist outside these values of the concentration, 1.5 and 0. It is also evident that very loose clusters evaporate before concentrated clusters, thus their stars do not have time to develop the BS characteristics. For the cluster to become old a value of rt/rc in between 1 and 12 is necessary, because the interaction of the cluster with the surrounding medium produces the evaporation of the cluster if this is not compact enough. Finally, old clusters become more and more compact, while loose clusters with rt/rc bigger than 12 (which in fact all are clusters without BS) are not favoured with a long life, in fact they disappear sharply for ages around log age=8.5.

Figure 4 displays how clusters, with and without BS stars, differ in concentration with respect to log age. This figure also illustrates that the concentration (c) is independent of the age and of the evaporation time but is in anticorrelation with the encounters (defined in the text) with slopes slightly different for the two samples; obviously, the more concentrated is the cluster, the higher is the number of encounters.

Once we know that clusters without BS dilute themselves in a shorter time than clusters with BS, the question is: are the BS stars sinking down indefinitely towards the centre of the cluster? In fact, stars, mainly the massive ones, are sinking down into the core, following a core collapse which is stopped at a certain time that we are going to determine below, providing that stars are not sinking down forever.

To confirm this statement, let us introduce a simple concept: the relation, in terms of log, between the core radius and the radius, rh, where half of the cluster mass is concentrated. We call this relation ch; ch = log (rc/rh).

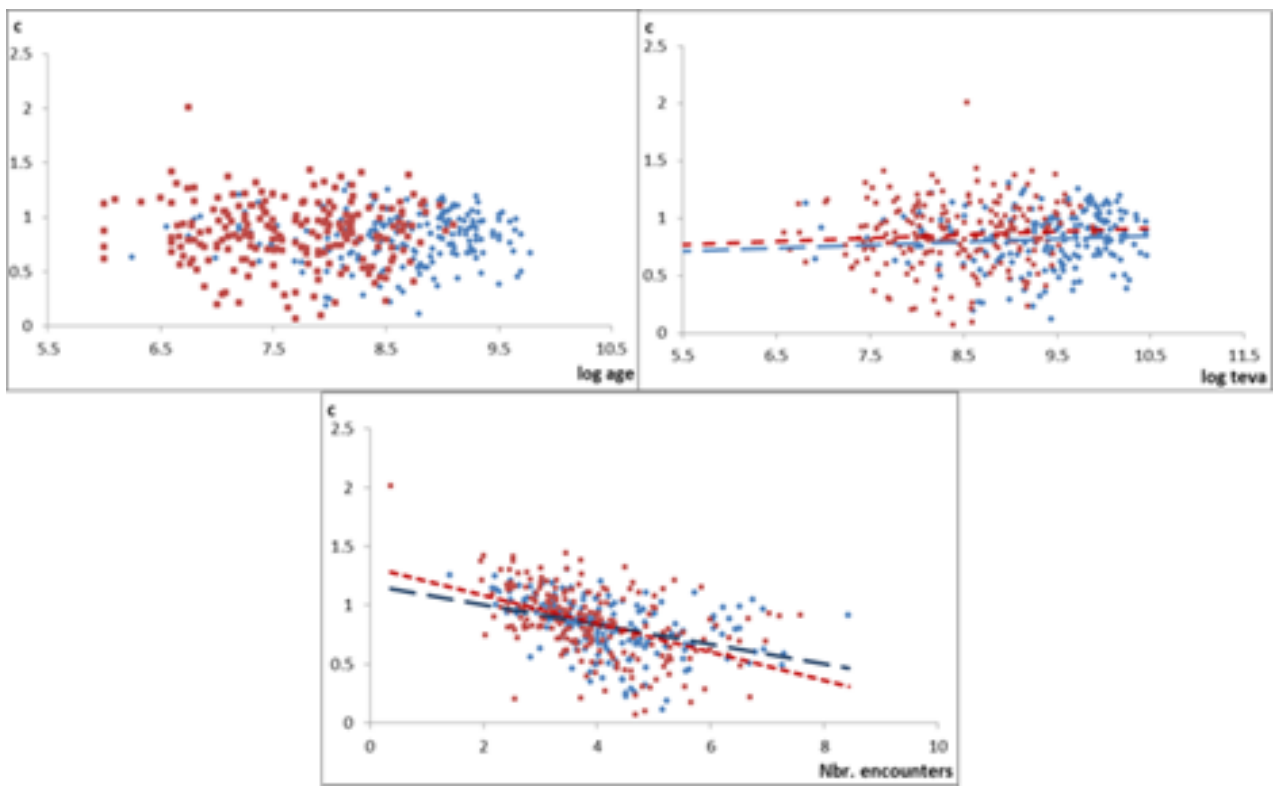

*Figure 4. Top-left relation between c and age; top-right: relation between c and evaporation time; bottom: relation between c and the number of encounters. Cluster NGC 2264 is represented by the isolated point.*

This kind of concentration, ch, opens a door in order to study the concentration towards the core collapse and where it stops. In fact, ch is an indicator of the cluster dynamical stage as Ferraro et al (2012) did with the BS stars distance to the cluster centre in core radius units. Here we use the rh in core radius units but with a different purpose. Let us compute equation (3) in the same way as it was explained in section 2.1; in which R →rh. We are able to derive from it the correlation between the half mass radius for clusters with BS stars and for clusters without BS stars and the timescale to reach rh; that means the timescale when the evolution of the cluster concentrates its mass in a volume whose radius is rh (th was listed in table 1). Results from these calculations are:

$$log\ rh\ (BS) = 1.12\ log\ rh\ (0\ BS) + 1.87$$
$$log\ th\ (BS) = 0.9\ log\ th\ (0\ BS) - 0.43$$

These two relations lead to a quick conclusion: in clusters containing BS stars, the volume which takes up half of the cluster mass is bigger than the one corresponding to clusters without BS stars. But the time to catch it up is shorter.

Based on GCs, Heggie and Hut (2003) found that a value for ch should be equal to – 3; however in our sample of OCs this value is much smaller. So that for OC with BS stars, ch varies between -0.7 and 0.4 (there are 28 over 190 clusters having ch ≥0) ; and for clusters without BS stars between -0.8 y 0.53 (there are 34 over 218 clusters with ch ≥0).

From all this discussion, we conclude that the core collapse has the same probability, to happen as well in clusters containing BS stars as in those that don't have them; in spite of their differences.

These last points give several question marks off, with respect to whether when the cluster evaporates the very dense core remains for longer; it is obvious that the denser is the core the higher is the gravity field intensity which can generate a gravitational collapse; however this core collapse should have to be stopped, or otherwise it would generate clusters with an very far as we know.

Let us analyse the last affirmation. In figure 5, is displayed ch versus age; this relation is showed to be constant; that means that there is no relevant trend, within the dispersion; however from log age ~9 and forward the ch for clusters with BS stars tends to 0 and therefore rc = rh.

According to Gieles et al (2008) the total mass of the cluster is Mcl α $r^3$ (r is the radius) then the concentration should be also a relation between the total mass of the cluster within the tidal radius and the cluster mass concentrated in the core; however equation (3) proves that this relation has to take into account both time and radius so that the "mass function" is a function of the radius and the time.

The $R^2$ = 0.9 (ch vs c) implies that this relation is linear, independently of the clusters containing or not containing BS stars but anti-correlated. We are able to argue the following: when c → 0.5 the ch → 0 the half mass radius and the core radius are the same; that means a half of the mass is concentrated in the core. However if c → 0.4, ch → 0.143, that means that rc is 40% larger than the rh; such could prove, if c → 0.4, that more than the half of the cluster mass is concentrated within the core.

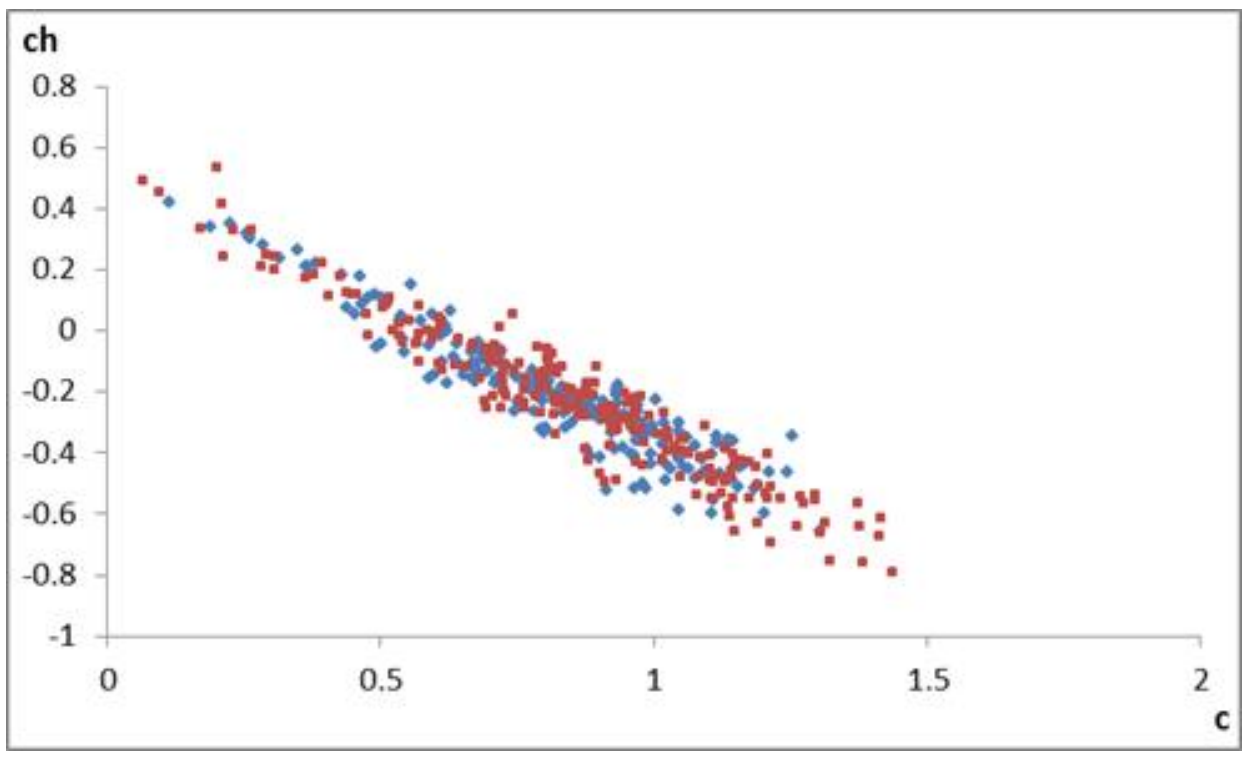

*Figure 5. In this figure it is showed that ch is independent of the age: contraction of the cluster could happen at any time. Apparently its behaviour is similar for those clusters with and without containing BS stars, but till log age ~9 after, that age the ch for clusters with BS stars tends to 0.*

If we compare figure 2 and figure 5, we find a direct relation between ch and the c; so that when ch tends to 0, then the tendency of c reaches a value of 0.5. Thus we can compare directly both "concentrations" as depicted in figure 6; as the best fit for the relation ch versus concentration could be written as:

$$ch = -0.8245\ c + 0.4732 \quad \textit{with an R2 = 0.9}$$

This conclusion suggests two reflexions: either the cluster collapses with the subsequent gravitational collapse and its consequences (blow up) or the collapse is stopped. In that last case, the core collapse is stopped because of the existence of a source of energy which sustains the collapse. We will be back to support this idea with the stars braking the collapse. It seems to us that ch, jointly with the c, is a good marker because it gives the depth of core collapse and the post collision core phase, as measured, for example, by the ratio of the core radius to half-mass radius rc/r h. If the relation rt/rc becomes larger, the relation rc/rh becomes smaller, let us look for the limit for the core collapse.

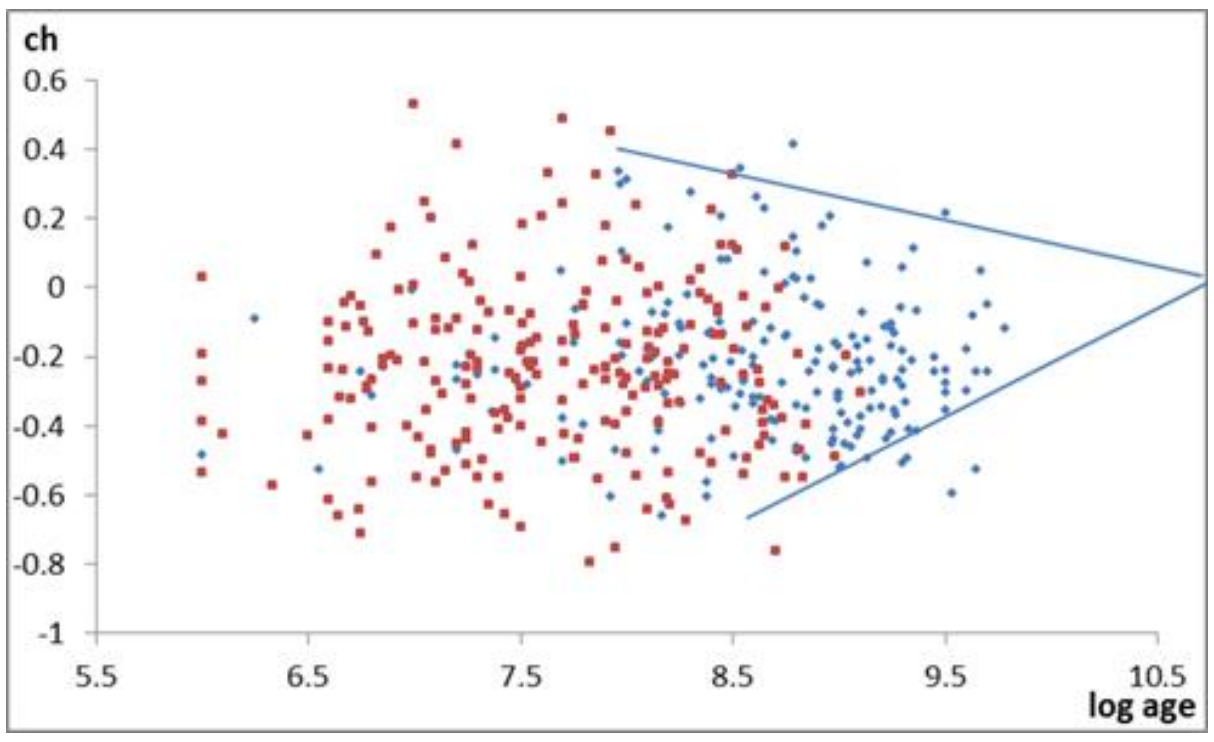

*Figure 6. The behaviour of ch is in anticorrelation with c.*

We investigate the behaviour of ch with respect to c, when rc ≥ rh, we found out that between the two values of c 0.4 and 0.5 corresponding to ch ~ 0.143 and 0, the relation presents a high dispersion, $R^2$ = 0.332 and $R^2$ = 0.43 corresponding to clusters with BS stars and without BS stars respectively; absolutely different from that got outside those values of c (0.4 and 0.5) as it can be seen in table 3. It could be interpreted as if the limit of the core collapse would be within a region bordered by rc= rh and rc ~1.40 rh. So that, such a border would indicate the existence of a region whose edges represent a transition zone where the equilibrium is broken; see values listed in table 3 which support the inflexion about the limit of the "core collapse". Let us propose ch as a "collapse marker" and their values of 0 and ~ 0.143 as the limit of a cocoon where the core collapse is stopped.

*2.3 Scheme of the model superimposed on real*

A scheme's rendering of the model super imposed on the plates of three example clusters (NGC 1912, NGC 6791 and NGC 2506) is found in Appendix A.

## 3 Expected number / presence of BS stars. (ƒ and ϖ)

### *3.1 Definition of f*

After a few crossings, the track of a star could be perturbed and might encounter another star; evidently interactions among stars are very important for dense clusters as happens in globular clusters but their probability is very low in the open clusters. In fact, we have found, in this work that the maximum number of encounters in OCs is limited, as already mentioned in section 2.1, to 8; which confirms the very low probability of produced encounters. Therefore the conclusion of this result is that the number of BS stars, created by star – star encounters during the life of the cluster is limited. This leads us to two possibilities: either BS stars were formed through primordial binaries since near the beginning or the formation of new BS stars occurs through encounters, throughout the cluster's life. Encounters could happen among isolated stars, binary stars or an isolated and a binary star.

*Table 2. Cluster near evaporation*

| Cluster Name | log age yr | Number BS | cross~ relax | log teva yr | Distan pc | Z kpc | RGC kpc |
|---|---|---|---|---|---|---|---|
| Haffner 18 | 7.9 | 0 | -0.1 | 8.8 | 23856 | 0.02 | 9.64 |
| NGC 2453 | 7.9 | 0 | 0.06 | 8.8 | 2383 | -0.04 | 9.32 |
| NGC 2467 | 8.1 | 0 | -0.1 | 8.6 | 1313 | 0.01 | 8.67 |
| Pismis 1 | 7.8 | 0 | 0.08 | 8.7 | 5903 | -0.07 | 11.1 |
| Trumpler 11 | 8.2 | 0 | -0.03 | 9 | 3884 | -0.33 | 7.98 |
| NGC 3324 | 6.1 | 0 | -0.05 | 7 | 2361 | -0.01 | 7.69 |
| Trumpler 14 | 6 | 0 | -0.05 | 6.6 | 2248 | -0.02 | 7.64 |
| Trumpler 16 | 6.5 | 0 | 0.03 | 7.5 | 2671 | -0.03 | 7.64 |
| Pismis 24 | 6 | 0 | 0.04 | 6.7 | 1481 | 0.02 | 6.53 |
| NGC 6396 | 7.5 | 0 | 0.04 | 8.2 | 11698 | -0.04 | 6.84 |
| NGC 6404 | 8.8 | 5 | -0.06 | 9.5 | 2146 | -0.04 | 5.86 |
| Ruprecht 130 | 8.5 | 0 | 0.03 | 9.3 | 3334 | -0.06 | 4.67 |
| NGC 6583 | 9 | 5 | -0.08 | 9.6 | 1753 | -0.08 | 6.28 |
| IC 5146 | 6 | 1 | -0.03 | 6.8 | 770 | -0.07 | 8.09 |

*Table 3. Linear anti-correlation between c and ch.*

| Clusters | ch | ch | $R^2$ |
|---|---|---|---|
| All clusters | | ch = - 0.83 c + 0.4695 | 0.90 |
| BS | < 0 | ch = - 0.72c + 0.3595 | 0.76 |
| BS | > 0.143 | ch = - 0.64 c + 0.4652 | 0.93 |
| 0 BS | < 0 | ch = - 0.75 c + 0.3953 | 0.82 |
| 0 BS | > 0.143 | ch = - 0.96 c + 0.5494 | 0.70 |
| **BS** | **0< ch<0.143** | **ch = - 0.31 c + 0.2226** | **0.33** |
| **0 BS** | **0< ch<0.143** | **ch = - 0.30 c + 0.2298** | **0.43** |

In bolt, the discontinuity found when 0< ch < 0.143 for both set of clusters: with and without BS stars.

Our results differ, in some aspects, from those obtained by Knigge (2015). We have found that the BS's number does correlate with dynamical encounters rate for young clusters. In the Knigge's work it was computed that he maximum number of encounters is 8. Such a number of encounters are too small for explaining the number of BS stars in some clusters, mainly in the old ones, then the origin by primordial binaries evolution is the preferred one in the case of old clusters and it can be added to the BS formed by encounters. We can confirm that, in old clusters the number of BS stars formed through direct collisions (dynamical BS stars) is negligible compared to the number of those formed through the evolution of primordial binaries (primordial BS stars); as it happens also in GCs (Davies et al 2004). But in young clusters the mass transfer in primordial binaries has not been enough to make these binaries to appear as BS stars. Therefore we think that the few BS stars present in our young clusters are of collisional origin.

Based on GCs, Mapelli et al. (2004, 2006) and Davies et al (2004) suggest scenarios combining BS from binary and collisional origin: different channels operate in different locations within a given cluster and different formation channels dominate in different clusters. Such an affirmation could explain the different behaviour but not the number of BS stars in clusters.

The aim of this section is to find out in which way BS stars are formed. Firstly, we define a ratio which might be equivalent to "frequency or probability of encounters and/or deflection of orbits" based upon the crossing time: Ƒ = age / tcross and $tcross = trlx/N/6 * ln(N/2)$. In order to have a normalized ratio we have defined a ratio, $ƒ \sim age/trlx$ , which is an indicator of the probability of encounter and/or deflection of orbits. This ratio ƒ is implicitly normalized to the total number of stars, N; thus ƒ is able to be compared among clusters, irrespective of their mass. The reason for our choice is based on the existence of a dynamical relaxation of stars trapped in the gravitational field and sunk down to the cluster core. The time to reach it can be estimated as the typical time that a star needs to change its energy by an amount equal to the main energy; this time was computed according to equation (3). As the time goes on, the effect extends larger and at larger distance; thus ƒ represents the number of deflections of the star due to encounters and/or due to the collective gravitational field created by the presence of previous stars; as a function of the age of the cluster. Secondly, we check and evaluate our definition and its relation with the number of BS stars observed, collected from the Ahumada and Lapasset (2007) catalogue; we count the number of BS within a bin of ƒ (cluster Berkeley 18 and its 126 BS stars has been taken off). Figure 7 shows the resultant histogram. If we connect the histogram discrete points with a line and smoothen it by means of a continuous line, thus we found that this line shows a curve whose shape exhibits an amazing result: the maximum of the total observed does not correspond to the maximum value of ƒ. Moreover the shape of the curve seems to be a distribution function, but such shape does not allow us to establish a simple relation between the number of BS and the age as was found by Leigh et al (2011) for GC.

Figure 8 shows that ƒ, which is the representative of the number of deflections it means the probability either of encounters or deflections of star trajectories or both, reaches values around 20 to 30 when the value of density is maximum. Also it reproduces the density (number of star per volume) vs ƒ: the maximum value of density is for ƒ values around 20 to 30. The shape of the envelope of this distribution, density with respect to ƒ, is alike as the one displayed in fig 7.

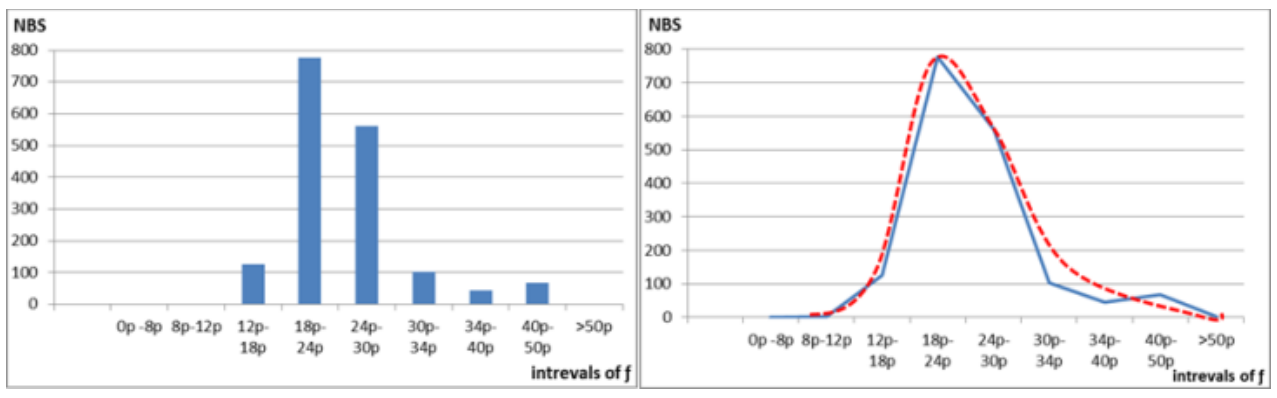

*Figure 7. It depictures a histogram representing the total number of "counted" number of BS stars within a bin of ƒ going from 0 to 50. The right drawing represents the shape of the histogram, connecting the top values, blue line and red discontinuous line. This envelope could represent the mathematical approach as described in the text.*

Figure 7 invites us to look for a mathematical expression which symbolizes the shape of the curve and has a physical significance. We found that this equation should be:

$$NBS \sim ƒ^3 \left(\frac{1}{e^{ƒ/\varpi} - 1}\right) \qquad \text{Equ (6)}$$

Where ƒ ~ age/trlx and $\varpi$ is a variable value related to the total number of BS stars, independently of their origin. One critical point would be the fact that the Number of BS stars (NBS) calculated here is not apparently normalized; but, the normalization is implicit – in ƒ - as it has been explained some paragraphs above.

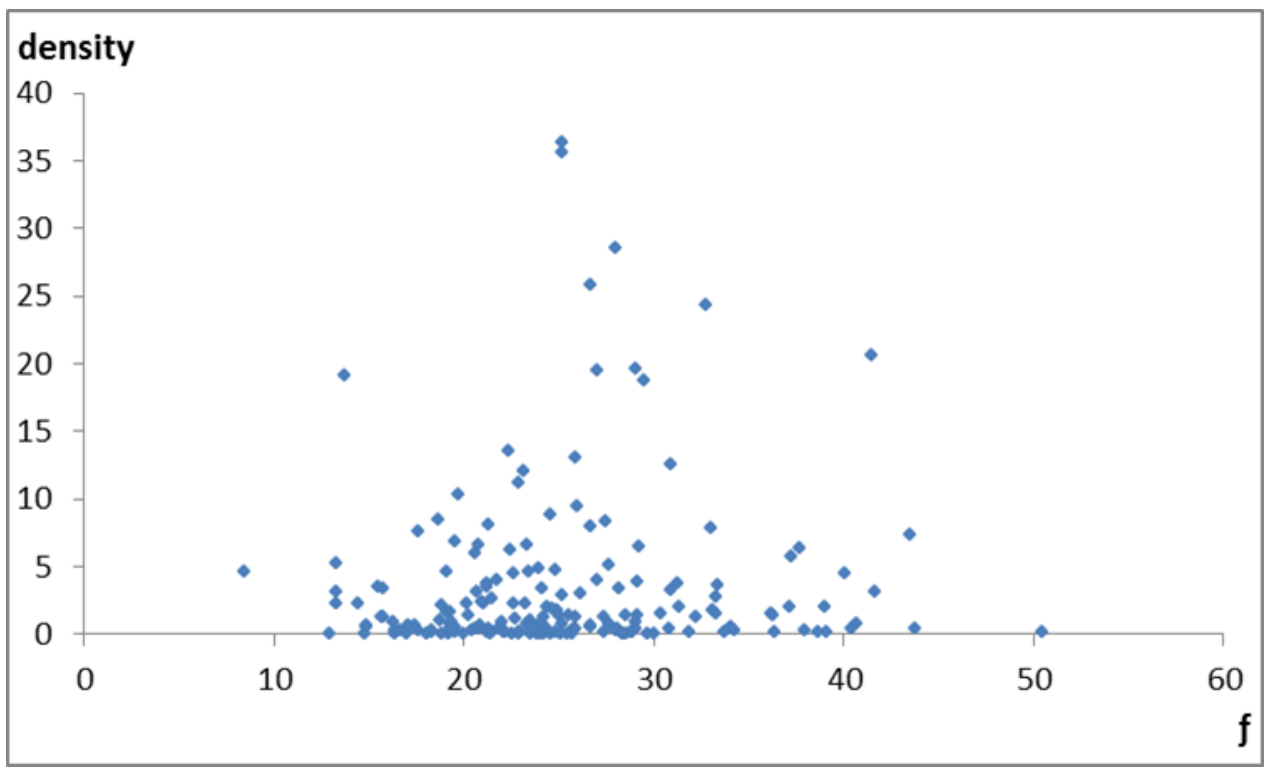

*Figure 8. It reproduces the density (number of stars per volume). Description is found in the text.*

The mathematical model has been represented for different values of the $\varpi$ in figure 9. It is very easy to see that figure 9 may overlap figure 8 to realise that the theoretical curves envelop on the observational data; so that larger number of BS in the cluster lager value of $\varpi$.

We verified whether equation 6 represents the number of BS stars present in a cluster; to do so, we checked by means of an empirical auditory. We have carried it out cluster by cluster, for all our sample clusters, showing that the number of observed BS stars which appears in an OC is not accidental, not random, but rather based on a law as expressed and formulated, by the equation (6). The graphic displayed in figure 10, certifies that the equation (6) is a reliable law representing the count of the number of BS stars within a cluster. Firstly, this figure 10 shows how good is the result of the theoretical calculation, based on the equation (6), and the observational values of the number of BS stars, per each intervals of ƒ. Secondly, the figure 10 convinces reliably and irrefutably that equation (6) represents the value of the number of BS contained in a cluster which has a determined value of age; the total number of the BS stars depends also on a factor ϖ which is an indicator of stellar collisions plus primordial binaries. The meaning and influence of ϖ is presented in the next section.

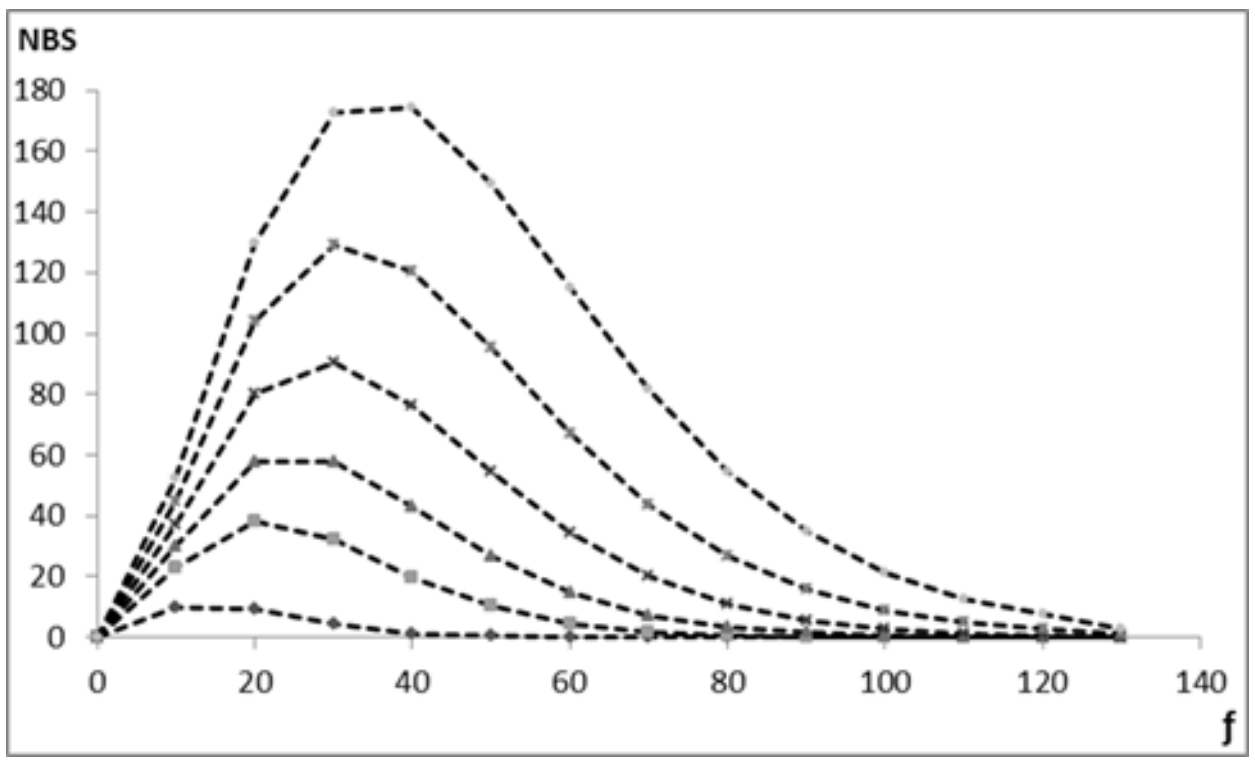

*Figure 9. It reproduces the values of equation 6 as function of f, computed for different values of ϖ: 2, 3, 3.5, 4, 4.5; respectively from the bottom to the top of the graphic.*

*3.2 Meaning of ϖ*

The explanation of the meaning of ϖ is not easy to do; let us go through its relation with respect to other variables, mainly encounters, age and concentration.

As showed in figure 9, the equation (6) is a faithful reflection of the number of BS. In table 4 we make a first approach regarding average values of ϖ, represented by < ϖ >, and number of encounters, as above defined. In table 5 are represented, by intervals of ϖ the observed BS stars in intervals, the average of the factor ϖ (<ϖ>) within each interval and the average of the encounters at the corresponding interval. It is easily derived that the number of encounters and ϖ do not correlate; however ϖ is the evidence of a combination of encounters and presence of the total number of BS stars.

However we want to study also how ϖ is related to the other variables. To do that we have built table 6 with the median values of the representative variables, within bins defined by the median of ϖ (let us represent it by [ϖ]).

A series of drawings were drawn to evaluate the dependence among the variables of the table 6. Therefore, figure 10 shows that there is a limit in the value of [ϖ] which cannot exceed a number around 5. Fig 10 also reproduces the relation between [ϖ] and c, as well as showing an inflexion point around log age ~9; such an age corresponds to the maximum value of number of BS stars in our sample.

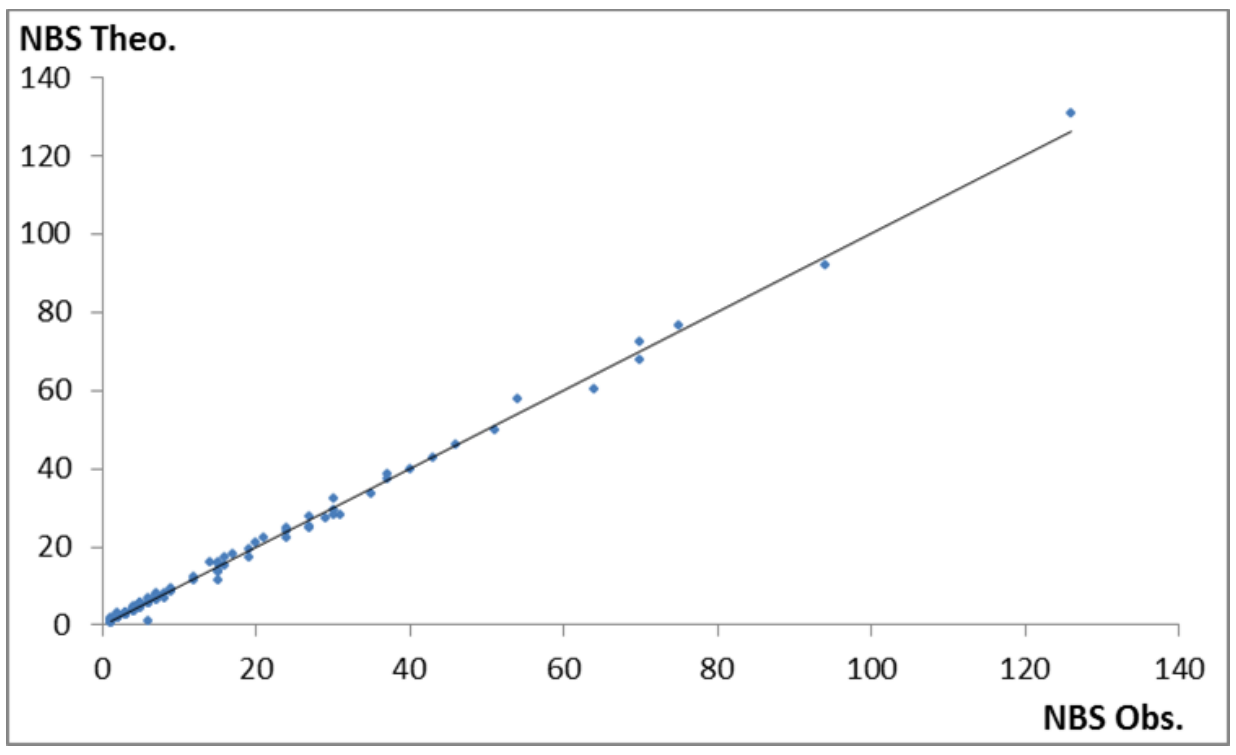

*Figure 10. This graphics showing that the number of BS stars computed from the theoretical/empirical equation 6 match with number of observed number of BS stars.*
*The relationship is linear with an $R^2$~ 1 and a slope by the same token ~ 1.*

From table 5, ch maintains its value between -0.36 and -0.17 (geometrical behaviour like a parabola) meanwhile concentration has values within a large range from 0.8 to 1.15. Table 5 also reveals that a discontinuity (or inflexion point) appears when [ϖ] is around 3.2 and [ch] has a value around 0.17 which is close to the value derived before ch = 0.143. Our interpretation is that during the process of evaporation, there is a drop of massive stars down to the core; such drop is stopped at the border of a region as described before. This phenomenon happens at log teva around 9.5 which correspond to log age ~ 9.0. At this age clusters, in our sample, without BS stars disappears.

*Table 4 Average value of ϖ facing to average number of encounters in intervals of number of BS stars.*

| Observed BS stars | <ϖ> | Average Encounters |
|---|---|---|
| 1 --5 | 2.7 | 4 |
| 5 -- 15 | 3.22 | 4 |
| 15 -- 25 | 3.47 | 3.7 ~ 4 |
| 25 –35 | 4.18 | 5 |
| 35 -- 45 | 4.18 | 4 |
| 45 -- 55 | 4.13 | 3.7 ~ 4 |
| 55-- 65 | 4.4 | 4 |
| 65 -- 75 | 4.6 | 4 |
| > 75 | 4.9 | 4 |

Finally, we have also applied the same equation (6) to another catalogue, as mentioned in the introduction, de Marchi et al. (2014), to show that our results are not due to any bias because of the selected catalogue.

To do that, we have computed the median values (values within []). Results are listed in table 6 in terms of the relationships between [ƒ], [ϖ] and [NBS]. The difference is, as expected, slight; evidently [ϖ] tends to 5. The better value of (R2) of de Marchi et al (2006) is due to the fact that their range of age is narrower than ours.

All previous discussion leads to the conclusion that ϖ has to be the indicator of stellar collisions plus primordial binary evolution related to the total number of observed BS stars.

*Table 5 Representative variables in intervals of median values of ϖ. [c] and [ch] are the median of c and ch for cluster with BS stars.*

| Median [ϖ] | ϖ adopt | [log age] yr | [log teva] yr | [ƒ] | [c] | [ch] |
|---|---|---|---|---|---|---|
| 1< ϖ <2 | 1.85 | 8.53 | 9.57 | 13.5 | 1.15 | -0.36 |
| 2≤ϖ <2.5 | 2.3 | 8.28 | 9.11 | 19.3 | 0.95 | -0.27 |
| 2.5≤ϖ <3 | 2.7 | 8.61 | 9.4 | 23.4 | 0.83 | -0.23 |
| 3≤ϖ <3.5 | 3.2 | 8.9 | 9.6 | 27.5 | 0.75 | -0.17 |
| 3.5≤ϖ <4 | 3.7 | 8.93 | 9.59 | 32.2 | 0.79 | -0.26 |
| 4≤ϖ <4.5 | 4.15 | 8.9 | 9.83 | 31.6 | 0.89 | -0.29 |
| 4.5≤ϖ <5 | 4.7 | 9.33 | 10.07 | 28.7 | 0.9 | -0.28 |
| 5≤ϖ <5.5 | 5.1 | 9.64 | 10.21 | 39.6 | 0.84 | -0.3 |

## 4. Summary and Conclusions

*3.1 Summary*

The clusters are not groups of static stars. They are dynamic since the very beginning. Their survival depends on the capacity of its gravitational energy to attract member stars. As time goes by the clusters tends to fade. We know that some star members go to the exterior after several millions of years, while others become concentrated. Lighter stars could escape and heavier ones go towards the centre.

Meanwhile there is a bustle of stars that follow trajectories that are deviated by encounters, attracted by the concentrated mass in the cluster core, and modified by certain proximity (joining stars forming binaries, etc.). Our first objective was to calculate the scale of those movements and compare them to deduce the residence times and escape times.

The first step was to assume that the mass of the cluster is the mass of its stars and that its variation at a specific point with time would be dM/dt, called it "function of mass". We elaborated this expression assuming that the density suffers minor variation (proven fact as from the moment the cluster starts to fade till it loses a significant part of itself, a long time passes).

Finally, after resolving the differential equation, the result is that the "function of mass" is proportional to r to the cube and to t to the square: $\chi \sim r^3\, t^2$. From here, the time – scale was calculated in the following way: if r corresponds with R (observed radius), then t is the age of the cluster; when r is rt then t is the evaporation time (teva); when r is rc then t is the collapsing time. The rest of the time-scales were calculated with the Armitage's equations.

These calculations allow us to derive the number of encounters, the dissolution time, and the conditions that the clusters comply before fading. These mentioned equations tell us that the number of encounters is small in comparison with the number of BS present in the case of old clusters, but is very similar for young ones.

Following the same philosophy, the time for concentrating half of the cluster mass (th) was calculated. We compared clusters with and without BS reaching the conclusion that the clusters with BS concentrate faster and occupy a bigger volume. The comparison of both values c = log (rt/rc) y ch = log(rc/rh) shows a lineal anticorrelation but with two points between which the anticorrelation is very poor; proof of the existence of a frontier, bordering two regions (like a cocoon). This border is where the core collapse is stopped

Once we calculated the relationship ƒ~ age/trlx, we drew a graphic NBS (Number of BS stars) vs ƒ, this relation would look like a distribution; this curve, empirically obtained, is reproduced by the mathematical expression: $NBS \sim ƒ^3 (\frac{1}{e^{ƒ/\varpi}-1})$ . Where ϖ is a factor whose values are between ~2 and ~5.

To validate this expression we compare the NBS calculated under the equation versus the NBS observed. The correlation is almost one to one; in fact:

$$NBS\ theo\ = 1.0043\ NBS\ obs\ -\ 0.0403\ with\ an\ R2 =\ 0.997.$$

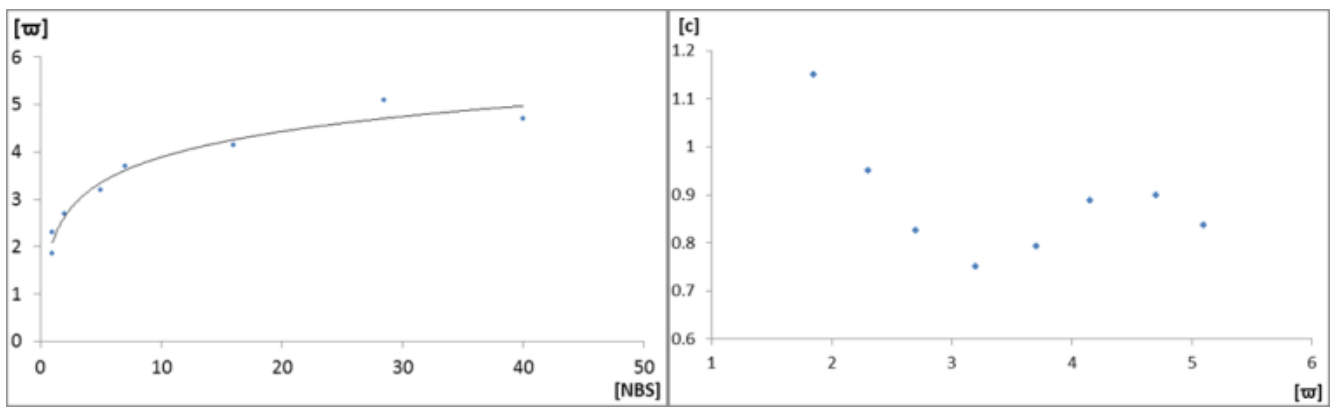

*Figure 10 Left: The behaviour of the median [ϖ] with respect to the median [Number of BS] is a logarithm function (see text), with an $R^2$ = 0.96; which points out the limitation of the [ϖ]. The value of [ϖ] tends to ~ 5, even for the cluster Berkeley 18 which contains the maximum number of BS stars - 126 BS stars - ϖ has the value of 5.1. Right: The behaviour of the median [c] with respect to the median [ϖ] is a polynomial function with an inflexion point at with at [ϖ] ~ 3.5 and ch ~ 0.75.*

*Table 6 Median values derived from the previous computation compared to those ones derived from de Marchi et al (2014)*

| | | |
|---|---|---|
| **Previous:** | | |
| [ϖ] vs [ƒ] | [ϖ] = 3.115 ln([ƒ])– 6.6 | $R^2$= 0.79 |
| [ϖ] vs [NBS] | [ϖ] = 0.782 ln([NBS])+2.1 | $R^2$= 0.96 |
| **de Marchi** | | |
| [ϖ] vs [ƒ] | [ϖ] = 3.500 ln ([ƒ]) – 7.7 | $R^2$= 0.93 |
| [ϖ] vs [NBS] | [ϖ] = 0.792 ln ([NBS]) +1.6 | $R^2$= 0.88 |

*4.2 Conclusions*

- Clusters not containing BS stars disappear younger than those clusters containing BS.
- The fraction of BS would "increase" due to dynamical effects.
- Major presence of BS stars, less difference exists in between age and evaporation, independently of the distance. Such a conclusion is common and well known, and corroborates the goodness of our model.
- We have determined that the number of encounters in OCs of log age ≤ 8 is in between 1 to 5 and in between 2 to 8 for those OCs whose age is log age >8.
- We have found that the common characteristics of clusters that find themselves near evaporation are: small Galactocentric distance, low amount of stars, young clusters, not containing BS
- Most of the clusters containing BS stars are within an interval of the c value (c=log (rt/rc)) between 0.4 and 1; which means that rt is in between 2.5 and 10 times the core radius.
- In clusters containing BS stars, the volume which takes up half the clusters mass is bigger than the one corresponding to clusters without BS stars. The time to reach the corresponding volume is shorter in cluster with BS stars than those ones without them.
- The relation of ch value (ch=log (rc/rh)) facing to c value is linear, independently of the clusters containing or not containing BS stars, but anti-correlated. We are able to argue the following: when $c \to 0.5$, $ch \to 0$ (the half mass radius and the core radius are the same). When $c \to 0.4$, $ch \to 0.143$, that means that rc is 1.4 times larger than the rh; such could prove that more than half of the cluster's mass is concentrated within the core.
- We have demonstrated through the relation between c and ch the existence of a region whose edges represent a transition zone where the linearity between them is broken; see values listed in table 3 which show the presence of an inflexion which represents the limit of the "core collapse". Let us propose ch as a "collapse marker" and their values of 0 and ~ 0.143 as the limits of a cocoon where the core collapse is stopped.
- The number/presence of BS is not random but follows a law which is expressed in terms of ƒ and ϖ as expressed by the equation (6):

$$NBS \sim f^3 \left(\frac{1}{e^{f/\varpi} - 1}\right)$$

- ƒ represents the number of deflections of the star due to encounters –originating merged stars - or due to the collective gravitational field created by the presence of previous stars, as a function of the age of the cluster. ϖ is the indicator of stellar collisions plus primordial binary presence. Its median value tends asymptotically to 5.

***Acknowledgements.***

Authors recognise the fruitful conversations and discussions with H. J. G. L. M. Lamers who provided us with many suggestions and recommendations, also gave important advices to us.

F. Llorente de Andrés acknowledges support from the Faculty of the European Space Astronomy Centre (ESAC). The reference number for the award was ESAC-391 dated on 17 May 2016.

F. Llorente de Andrés recognises the influence of Garcillán (Segovia) pannier maker people in trying to look beyond the sky.

C. Morales Durán acknowledged partial support from the Spanish Virtual Observatory (http://svo.cab.inta-csic.es) supported from the Spanish MINECO through grants AyA2011-24052 and AYA2014-55216.

## Appendix A

*Scheme and cluster picture are rendering of the model as described in the section*

The figure shows the scheme of the model described in Section 2 and it is superimposed on the photo/cluster chart of three OCs. In the scheme r2 corresponds to cluster radius, rt corresponds to the tidal radius; rh the radius where half of the mass is concentrated.

It is shown how changes of tracks (light blue colour) could force the star to encounter another one and both stars could merge in one star or in a pair of closed binaries; travelling jointly together. It is also easy to realise that due to dynamical effects stars are removed from the outskirts (mix red and dark blue colour) most probably the low mass stars. Thus, that could be interpreted as that relaxation will force the massive stars to the central regions, and could force the lower mass stars to the periphery.

Moreover, during the core collapse, the massive stars attempt to equalize kinetic energy with stars outside the core; they lose energy and sink even faster toward the core. The border of this process limit is called core collapse; in the picture this border is identified by an internal circle whose radius is rc. That is described in detail in section 2; where also is discussed that that means that rc could be 40% larger than the rh if concentration c tends to 0.4, that means that more than the half of the cluster mass is concentrated within the core. The picture of NGC 6791 is a good example.

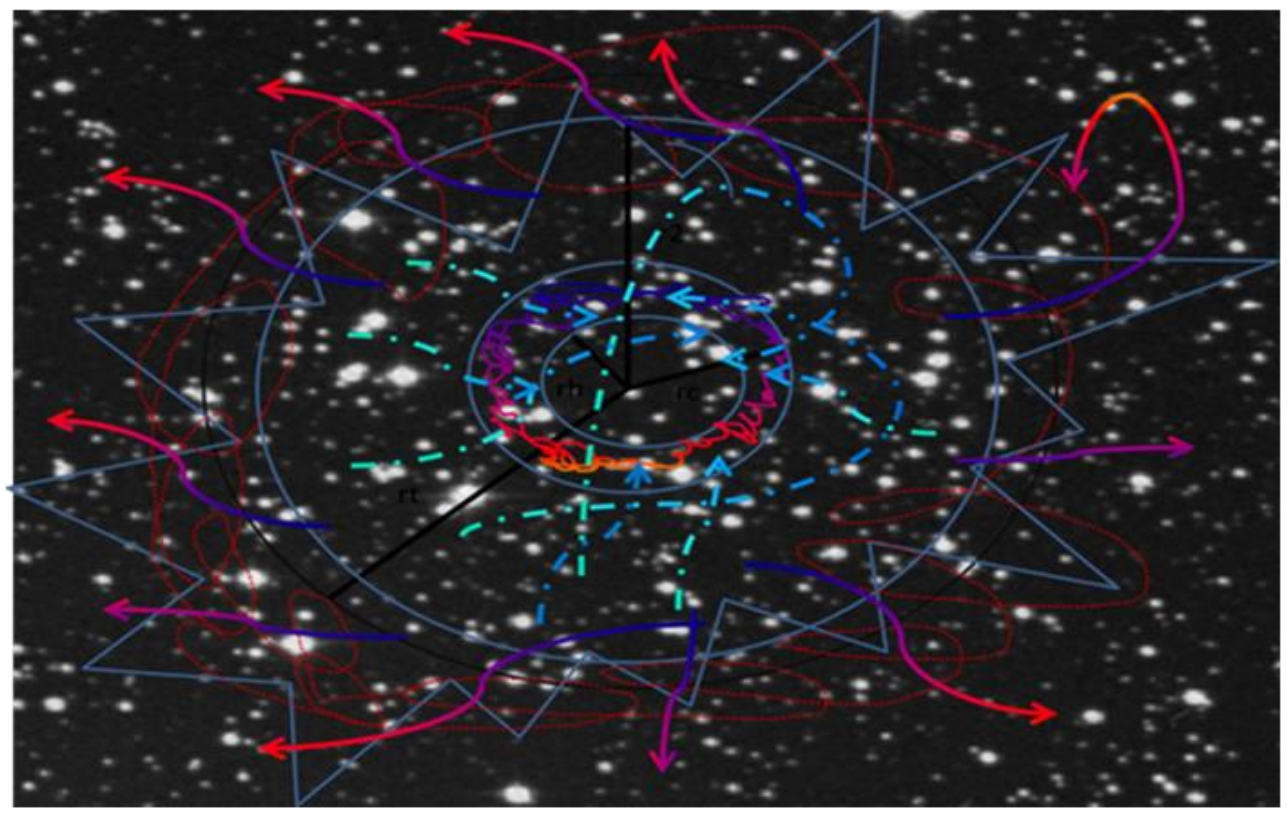

NGC 1912; log age = 8.35; 0 BS; log teva 8.54; ƒ = 31.84; ϖ = 0; Cocoon = -0..02

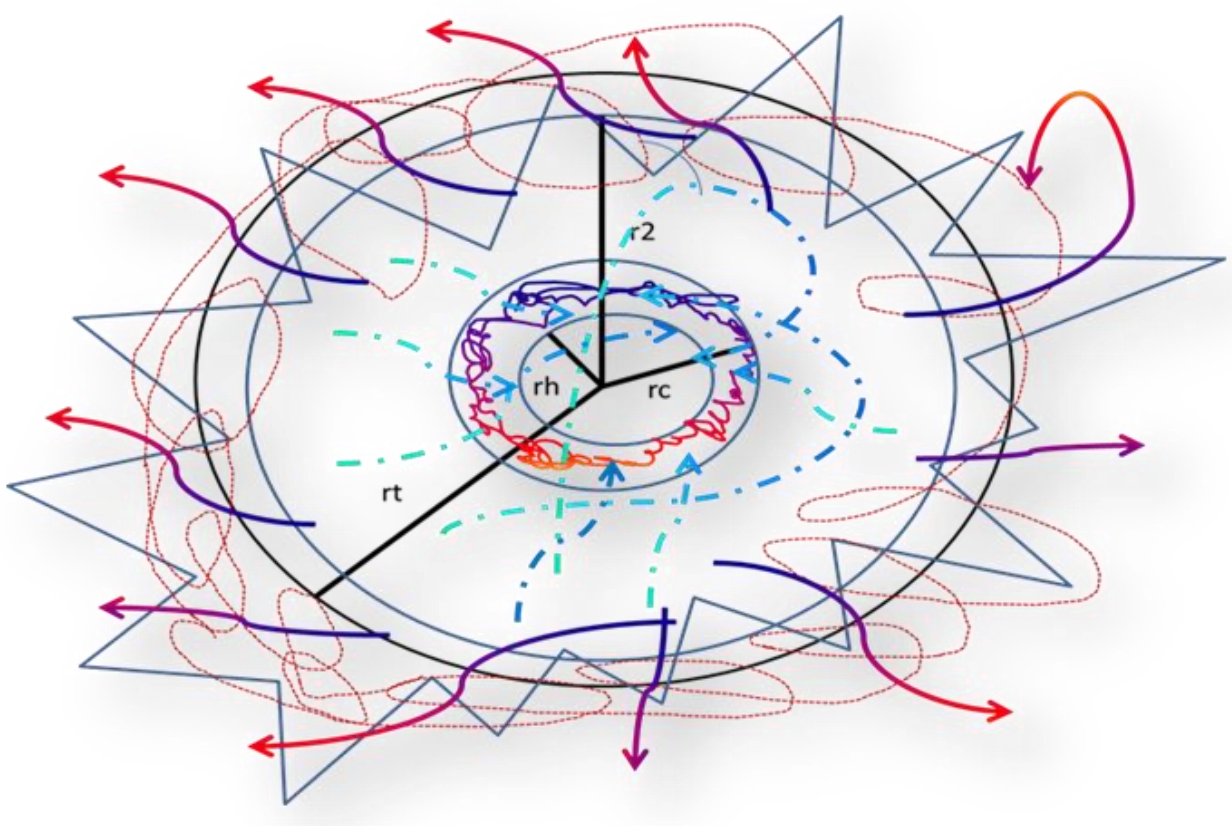

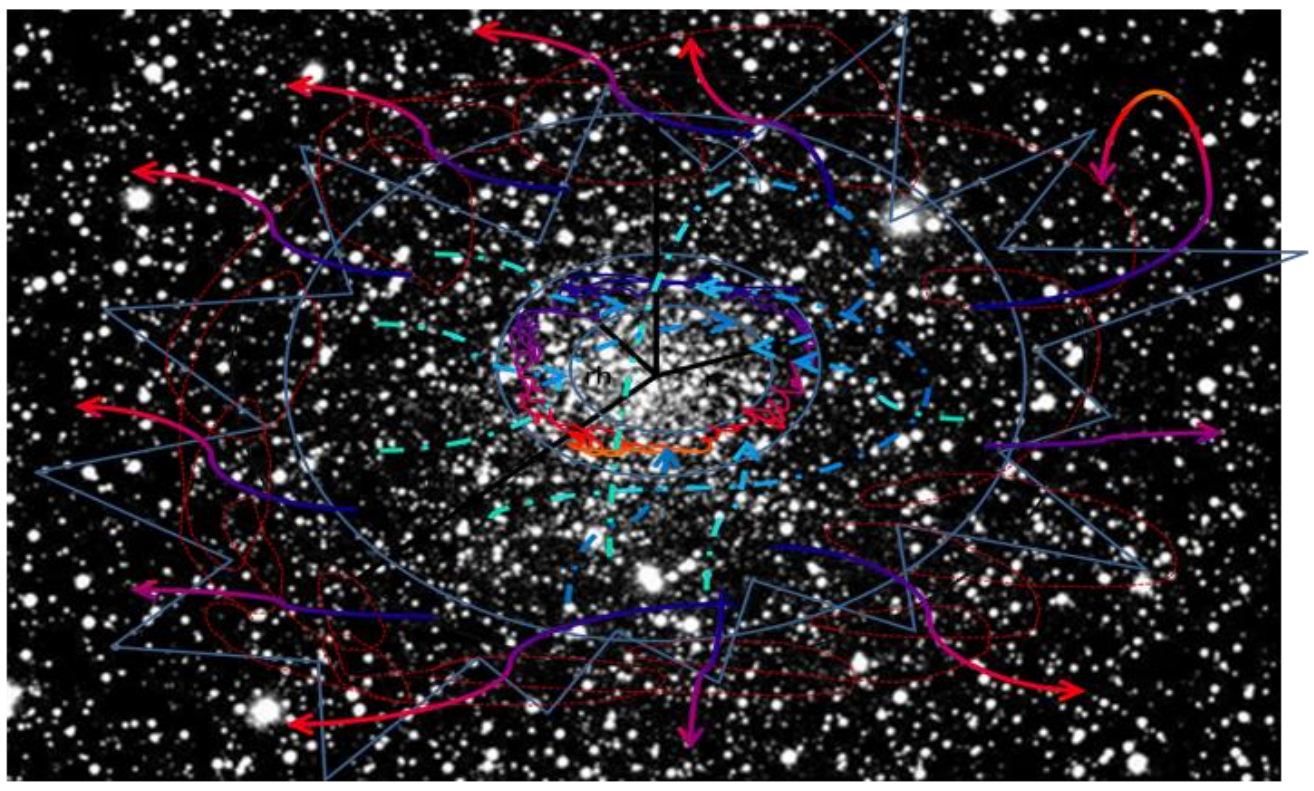

NGC 6791; log age = 9.645; 75 BS; log teva 10.45; ƒ = 21.43; ϖ = 4.4; Cocoon = -0.24

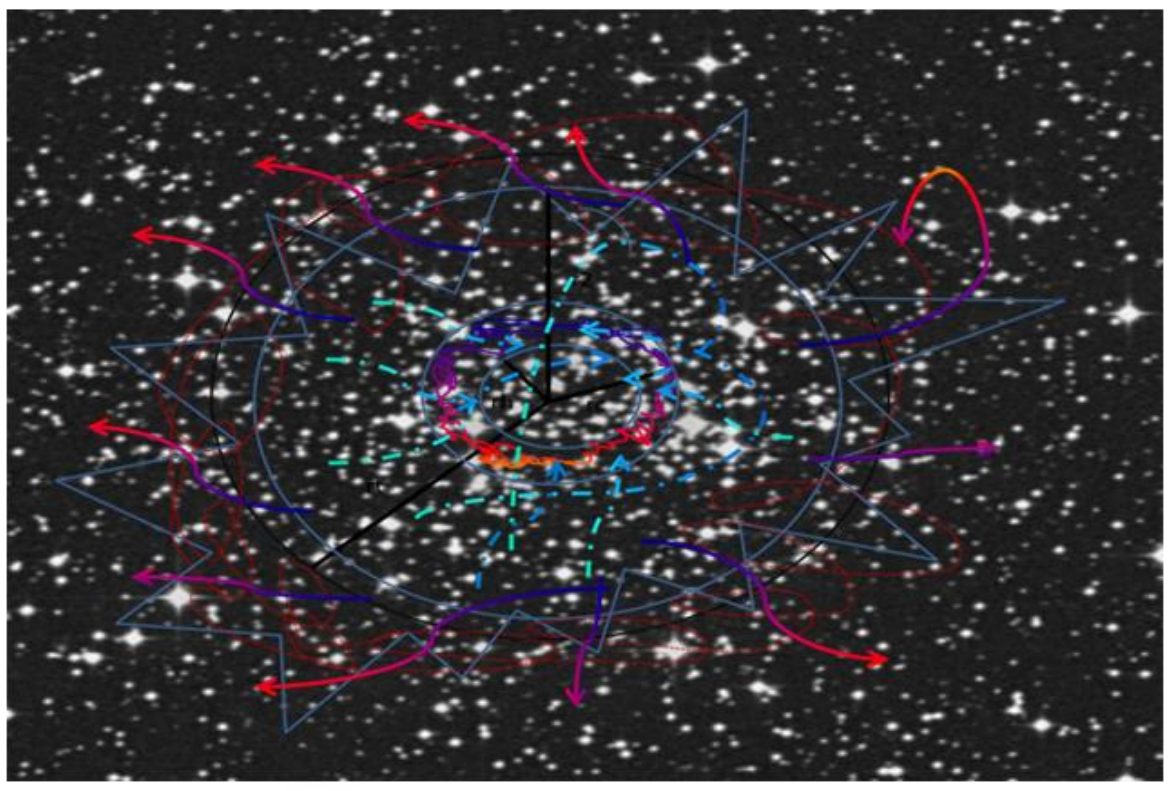

NGC 2506; log age = 9.21; 15 BS; log teva 9.93; ƒ = 25.67; ϖ = 3.6; Cocoon = -0.26

**Appendix B**

List of our sample of Open Clusters with their respective ƒ and adopted ϖ; including the observed BS stars and those predicted by our model. Remark that almost all clusters (85% of the sample) show a value of ϖ between 2 and 4.

| Cluster Name | ƒ Max | NBS Obs | NBS Theo | ϖ Adopt |
|---|---|---|---|---|
| Berkeley 104 | 20.6 | 9 | 9 | 3 |
| Berkeley 12 | 19.29 | 15 | 12 | 3 |
| Berkeley 14 | 22.98 | 94 | 92 | 4.7 |
| Berkeley 17 | 25.57 | 31 | 28 | 3 |
| Berkeley 18 | 23.5 | 126 | 131 | 5.1 |
| Berkeley 19 | 19.33 | 1 | 1 | 2.2 |
| Berkeley 20 | 28.48 | 8 | 7 | 3.5 |
| Berkeley 21 | 18.32 | 51 | 50 | 3.8 |
| Berkeley 22 | 18.82 | 21 | 22 | 3.3 |
| Berkeley 23 | 16.97 | 12 | 11 | 2.8 |
| Berkeley 28 | 21.07 | 1 | 1 | 2.3 |
| Berkeley 29 | 14.77 | 7 | 7 | 2.4 |
| Berkeley 30 | 16.34 | 1 | 1 | 2 |
| Berkeley 31 | 18 | 20 | 21 | 3.2 |
| Berkeley 32 | 22.14 | 37 | 37 | 3.9 |
| Berkeley 33 | 30.05 | 2 | 2 | 3.1 |
| Berkeley 39 | 24.57 | 43 | 43 | 4.2 |
| Berkeley 60 | 15.72 | 2 | 2 | 2.1 |
| Berkeley 64 | 16.33 | 3 | 3 | 2.2 |
| Berkeley 66 | 21.47 | 70 | 68 | 4.3 |
| Berkeley 69 | 21.29 | 2 | 2 | 2.5 |
| Berkeley 70 | 23.85 | 64 | 60 | 4.4 |
| Blanco 1 | 43.73 | 1 | 1 | 3.8 |
| Collinder 110 | 28.08 | 27 | 28 | 4.2 |
| Collinder 74 | 33.99 | 7 | 6 | 3.9 |
| Haffner 17 | 20.83 | 1 | 1 | 2.3 |
| Haffner 6 | 17.63 | 2 | 2 | 2.2 |
| IC 166 | 21.24 | 7 | 8 | 3 |
| King 15 | 19.21 | 3 | 3 | 2.5 |
| King 2 | 33.72 | 30 | 29 | 4.7 |
| King 5 | 21.98 | 4 | 4 | 2.8 |
| King 6 | 25.22 | 1 | 1 | 2.7 |
| King 7 | 30.79 | 6 | 6 | 3.6 |
| King 8 | 19.3 | 5 | 6 | 2.7 |
| Melotte 20 | 38.61 | 1 | 1 | 3.5 |
| Melotte 66 | 24.19 | 35 | 34 | 4 |
| Melotte 71 | 28.25 | 3 | 3 | 3.2 |
| NGC 1193 | 24.07 | 19 | 17 | 3.6 |
| NGC 1245 | 23.84 | 7 | 8 | 3.2 |
| NGC 1252 | 27.63 | 1 | 1 | 2.7 |
| NGC 1664 | 28.14 | 1 | 1 | 2.8 |
| NGC 1798 | 24.82 | 24 | 22 | 3.8 |
| NGC 1817 | 24.95 | 5 | 5 | 3.1 |
| NGC 188 | 39.05 | 24 | 24 | 5 |
| NGC 1893 | 50.49 | 2 | 2 | 4.5 |
| NGC 1901 | 13.71 | 1 | 2 | 1.9 |
| NGC 1907 | 21.72 | 6 | 6 | 2.9 |
| NGC 2099 | 33.07 | 3 | 3 | 3.5 |
| NGC 2112 | 16.3 | 15 | 16 | 2.9 |
| NGC 2126 | 22.58 | 1 | 1 | 2.5 |
| NGC 2141 | 22.87 | 24 | 25 | 3.7 |
| NGC 2158 | 28.74 | 40 | 40 | 4.5 |
| NGC 2168 | 39.02 | 1 | 1 | 3.5 |
| NGC 2192 | 36.32 | 3 | 3 | 3.7 |
| NGC 2194 | 25.77 | 15 | 16 | 3.7 |
| NGC 2204 | 22.13 | 9 | 9 | 3.1 |
| NGC 2243 | 22.91 | 9 | 9 | 3.2 |
| NGC 2251 | 20.63 | 3 | 3 | 2.6 |
| NGC 2259 | 18.26 | 4 | 4 | 2.5 |
| NGC 2266 | 16.66 | 2 | 2 | 2.2 |
| NGC 2281 | 27.5 | 1 | 1 | 2.7 |
| NGC 2287 | 33.26 | 2 | 2 | 3.4 |
| NGC 2323 | 31.26 | 2 | 2 | 3.3 |
| NGC 2324 | 29.67 | 4 | 4 | 3.4 |
| NGC 2353 | 20.19 | 1 | 1 | 2.3 |
| NGC 2354 | 23.31 | 3 | 3 | 2.8 |
| NGC 2355 | 29.02 | 5 | 5 | 3.4 |
| NGC 2360 | 24.68 | 4 | 4 | 3 |
| NGC 2383 | 15.75 | 1 | 1 | 1.9 |
| NGC 2420 | 28.83 | 5 | 5 | 3.4 |
| NGC 2422 | 37.71 | 1 | 1 | 3.5 |
| NGC 2437 | 32.23 | 7 | 7 | 3.8 |
| NGC 2439 | 22.56 | 1 | 1 | 2.4 |
| NGC 2447 | 23.25 | 3 | 3 | 2.8 |
| NGC 2451 | 43.51 | 2 | 2 | 4.1 |
| NGC 2477 | 30.37 | 15 | 14 | 4 |
| NGC 2506 | 25.67 | 15 | 14 | 3.6 |
| NGC 2516 | 22.36 | 1 | 1 | 2.4 |
| NGC 2527 | 30.86 | 1 | 1 | 3 |
| NGC 2533 | 20.74 | 1 | 1 | 2.3 |

| | | | | | | | | | |
|---|---|---|---|---|---|---|---|---|---|
| NGC 129 | 23.92 | 1 | 1 | 2.6 | NGC 2539 | 23.54 | 1 | 1 | 2.5 |
| NGC 1342 | 22.86 | 2 | 2 | 2.6 | NGC 2546 | 13.25 | 2 | 2 | 1.9 |
| NGC 1348 | 29.1 | 1 | 1 | 2.8 | NGC 436 | 34.23 | 2 | 2 | 3.5 |
| NGC 1444 | 13.32 | 2 | 3 | 2 | NGC 457 | 40.63 | 1 | 1 | 3.7 |
| NGC 559 | 26.68 | 1 | 1 | 2.7 | NGC 5822 | 33.03 | 7 | 8 | 3.9 |
| NGC 744 | 24.37 | 1 | 1 | 2.6 | NGC 5823 | 26.69 | 1 | 1 | 2.7 |
| NGC 752 | 34.1 | 1 | 1 | 3.3 | NGC 5999 | 15.49 | 4 | 4 | 2.3 |
| NGC 884 | 14.85 | 2 | 2 | 2 | NGC 6005 | 15.68 | 16 | 17 | 2.9 |
| Pleiades | 37.97 | 1 | 1 | 3.5 | NGC 6025 | 19.73 | 1 | 1 | 2.2 |
| Stock 2 | 24.03 | 3 | 3 | 2.8 | Ruprecht 115 | 33.31 | 2 | 2 | 3.4 |
| Tombaugh 1 | 23.99 | 4 | 5 | 3 | NGC 6067 | 28.55 | 3 | 3 | 3.2 |
| Tombaugh 2 | 12.91 | 17 | 18 | 2.7 | NGC 6124 | 25.2 | 1 | 1 | 2.6 |
| Trumpler 5 | 28.03 | 70 | 72 | 4.9 | NGC 6134 | 41.47 | 6 | 6 | 4.4 |
| Trumpler 9 | 36.22 | 1 | 1 | 3.4 | NGC 6253 | 40.07 | 27 | 25 | 5.1 |
| Pismis 2 | 23.95 | 46 | 46 | 4.2 | NGC 6259 | 25.19 | 6 | 6 | 3.2 |
| Collinder 185 | 24.81 | 1 | 1 | 2.6 | IC 4651 | 23.12 | 6 | 6 | 3 |
| Pismis 3 | 23.32 | 30 | 32 | 3.9 | NGC 6404 | 21.81 | 5 | 6 | 2.9 |
| NGC 2627 | 25.86 | 14 | 16 | 3.7 | NGC 6416 | 21.47 | 1 | 1 | 2.4 |
| Pismis 7 | 28.38 | 3 | 3 | 3.2 | NGC 6425 | 14.46 | 1 | 1 | 1.8 |
| Ruprecht 67 | 19.53 | 1 | 1 | 2.2 | NGC 6451 | 20 | 3 | 3 | 2.5 |
| NGC 2660 | 17.43 | 8 | 8 | 2.7 | NGC 6494 | 8.44 | 2 | 2 | 1.5 |
| NGC 2658 | 25.45 | 4 | 3 | 3 | NGC 6583 | 36.3 | 5 | 5 | 4 |
| NGC 2669 | 29.23 | 1 | 1 | 2.9 | NGC 6603 | 22.03 | 3 | 3 | 2.7 |
| Trumpler 10 | 32.71 | 1 | 1 | 3.1 | NGC 6631 | 16.95 | 3 | 3 | 2.3 |
| NGC 2682 | 40.37 | 30 | 28 | 5.2 | NGC 6633 | 29.46 | 4 | 4 | 3.4 |
| NGC 2818 | 23.84 | 2 | 2 | 2.7 | IC 4725 | 27.97 | 5 | 5 | 3.3 |
| NGC 2849 | 31.86 | 4 | 4 | 3.5 | NGC 6649 | 41.62 | 1 | 1 | 3.8 |
| Ruprecht 75 | 19.52 | 9 | 9 | 2.9 | IC 4756 | 30.88 | 6 | 1 | 3 |
| vdB-Hagen 66 | 27.35 | 2 | 2 | 3 | Berkeley 79 | 19.29 | 1 | 1 | 2.2 |
| IC 2488 | 26.67 | 3 | 3 | 3 | NGC 6694 | 18.81 | 1 | 1 | 2.2 |
| Ruprecht 79 | 20.92 | 1 | 1 | 2.3 | NGC 6705 | 25.19 | 1 | 1 | 2.6 |
| NGC 3114 | 27.6 | 5 | 5 | 3.3 | NGC 6709 | 17.56 | 2 | 3 | 2.3 |
| Collinder 223 | 21.32 | 2 | 2 | 2.5 | Berkeley 81 | 14.88 | 19 | 20 | 2.9 |
| NGC 3293 | 25.89 | 1 | 1 | 2.6 | NGC 6755 | 24.96 | 1 | 1 | 2.6 |
| NGC 3496 | 21.24 | 1 | 1 | 2.3 | NGC 6791 | 21.35 | 75 | 77 | 4.4 |
| NGC 3532 | 29.07 | 5 | 5 | 3.4 | Stock 1 | 27.02 | 2 | 2 | 2.9 |
| Trumpler 18 | 20.77 | 1 | 1 | 2.3 | NGC 6811 | 24.14 | 1 | 1 | 2.5 |
| Stock 13 | 13.28 | 1 | 1 | 1.7 | NGC 6819 | 23.39 | 29 | 27 | 3.8 |
| IC 2714 | 25.85 | 2 | 2 | 2.9 | NGC 6834 | 23.42 | 3 | 3 | 2.8 |
| Melotte 105 | 29.12 | 4 | 4 | 3.3 | Roslund 3 | 37.28 | 2 | 2 | 3.7 |
| NGC 3680 | 22.4 | 1 | 1 | 2.4 | NGC 6871 | 22.71 | 2 | 2 | 2.6 |
| NGC 3766 | 22.6 | 1 | 1 | 2.4 | IC 1311 | 21.41 | 12 | 12 | 3.2 |
| IC 2944 | 29.08 | 1 | 1 | 2.9 | NGC 6939 | 24.22 | 5 | 4 | 3 |
| NGC 3960 | 20.22 | 1 | 1 | 2.3 | NGC 6940 | 33.35 | 6 | 6 | 3.8 |
| Ruprecht 98 | 25.93 | 1 | 1 | 2.7 | IC 1369 | 19.32 | 6 | 6 | 2.7 |
| NGC 4349 | 19.06 | 1 | 1 | 2.2 | NGC 7044 | 19.57 | 6 | 7 | 2.8 |

| | | | | |
|---|---|---|---|---|
| Collinder 261 | 24.27 | 54 | 58 | 4.4 |
| NGC 4755 | 23.68 | 1 | 1 | 2.5 |
| NGC 4815 | 19.24 | 5 | 6 | 2.7 |
| NGC 5168 | 26.14 | 1 | 1 | 2.7 |
| Pismis 18 | 18.74 | 3 | 3 | 2.4 |
| Lynga 1 | 18.66 | 1 | 1 | 2.2 |
| NGC 7510 | 21.22 | 1 | 1 | 2.4 |
| Berkeley 99 | 25.06 | 7 | 6 | 3.2 |
| King 11 | 20.4 | 27 | 25 | 3.5 |
| NGC 7082 | 27.02 | 1 | 1 | 2.8 |
| NGC 7086 | 24.55 | 5 | 5 | 3.1 |
| NGC 7142 | 26.67 | 37 | 39 | 4.3 |
| IC 5146 | 20.69 | 1 | 1 | 2.3 |
| NGC 7160 | 19.12 | 1 | 1 | 2.2 |
| NGC 7243 | 31.34 | 1 | 1 | 3.1 |
| NGC 7788 | 37.17 | 3 | 3 | 3.8 |
| NGC 7789 | 27.4 | 16 | 15 | 3.8 |
| NGC 7790 | 17.08 | 3 | 3 | 2.3 |